\documentclass[journal]{IEEEtran}
\IEEEsettopmargin{t}{54pt}
\usepackage{epsfig,amsmath,amssymb,epsf,amsthm,scalefnt,multirow}
\usepackage{xcolor}
\usepackage{float}
\usepackage{cite}
\usepackage{psfrag}
\usepackage{bm}
\usepackage{algorithmic,algorithm}
\usepackage{graphicx}
\usepackage{mleftright}
\DeclareGraphicsExtensions{.pdf,.png,.jpg}
\usepackage{subfigure}

\newtheorem{theorem}{Theorem}
\newtheorem{lemma}{Lemma}
\newtheorem*{lemma*}{Lemma}

\newcommand{\norm}[1]{\left\lVert#1\right\rVert}

\usepackage[labelformat=simple,font={footnotesize,sf}]{subcaption}

\begin{document}
\title{Coverage Analysis of Dynamic Coordinated Beamforming for LEO Satellite Downlink Networks}

\author{Daeun~Kim,~\IEEEmembership{Student~Member,~IEEE},~Jeonghun Park,~\IEEEmembership{Member,~IEEE},~and~Namyoon~Lee,~\IEEEmembership{Senior~Member,~IEEE}

\thanks{D. Kim is with the Department of Electrical Engineering, POSTECH, Pohang, Gyeongbuk 37673, South Korea  (e-mail: daeun.kim@postech.ac.kr).}
\thanks{J. Park is with the School of Electrical and Electronic Engineering, Yonsei University, Seoul 02841, South Korea (email:
jhpark@yonsei.ac.kr).}
	\thanks{N. Lee is with the School of Electrical Engineering, Korea University, Seoul 02841, South Korea (e-mail: namyoon@korea.ac.kr).}
}

  \maketitle

\begin{abstract}
In this paper, we investigate the coverage performance of downlink satellite networks employing dynamic coordinated beamforming. Our approach involves modeling the spatial arrangement of satellites and users using Poisson point processes situated on concentric spheres. We derive analytical expressions for the coverage probability, which take into account the in-cluster geometry of the coordinated satellite set. These expressions are formulated in terms of various parameters, including the number of antennas per satellite, satellite density, fading characteristics, and path-loss exponent. To offer a more intuitive understanding, we also develop an approximation for the coverage probability. Furthermore, by considering the distribution of normalized distances, we derive the spatially averaged coverage probability, thereby validating the advantages of coordinated beamforming from a spatial average perspective. Our primary finding is that dynamic coordinated beamforming significantly improves coverage compared to the absence of satellite coordination, in direct proportion to the number of antennas on each satellite. Moreover, we observe that the optimal cluster size, which maximizes the ergodic spectral efficiency, increases with higher satellite density, provided that the number of antennas on the satellites is sufficiently large. Our findings are corroborated by simulation results, confirming the accuracy of the derived expressions.

\end{abstract}


\section{Introduction}
With the growing demand for ubiquitous connectivity, particularly in the context of the future 6G wireless network, the need for extensive coverage becomes essential. Satellite networks are regarded as an effective solution to provide seamless connectivity across the globe, serving diverse applications \cite{Zhu2022,Liu2021}. Especially, low Earth orbit (LEO) satellite networks have garnered significant attention due to their low propagation delay and signal attenuation, offering improved data rates and low latency compared to traditional geostationary Earth orbit (GEO) satellite networks. Recently, several LEO satellite projects, such as Starlink and OneWeb, have proposed to provide global access to the Internet with thousands of satellites \cite{Del2019}.

However, many LEO satellites are required to achieve seamless coverage across the entire Earth's surface in LEO satellite networks. The substantial number of satellites nearby increases interference among neighboring satellites. Further, in satellite networks, the coverage probability decreases significantly with increasing network density \cite{Park2023}. In the cellular network, however, the probability of coverage does not depend on the base station (BS) density \cite{Andrews2011}. This discrepancy shows that the satellite networks are notably and adversely affected by interference. Consequently, addressing interference mitigation becomes a critical imperative in satellite network design and operation \cite{Su2019}. One of the promising solutions is satellite coordination. In cellular networks, BS coordination is regarded as a practical approach to mitigate inter-cell interference \cite{Clerckx2013,Lee2012}. Among various BS coordination approaches, dynamic clustering-based BS coordination offers high spectral efficiency gains by creating a coordination set with BSs that has substantial interference power \cite{Lee2015-2}. By doing this, interference coming from out-of-cluster BSs is effectively mitigated.

Understanding the coverage and rate performance of LEO satellite networks, with satellite coordination, is essential for the initial design and planning of satellite networks. Moreover, comprehending coverage allows network operators to make cost-effective decisions. The conventional Walker constellation that places the satellites on a grid has been used to characterize the satellite network performance. However, this approach requires complex system-level simulations to marginalize various sources of randomness. Further, it is not clear how to incorporate the interference mitigation technique such as satellite coordination. Therefore, there is a necessity for an analytical tool to characterize the coverage performance of satellite networks with satellite coordination, facilitating insights for network design. In this work, we analyze the coverage performance of downlink satellite networks considering satellite coordination, with a particular focus on dynamic coordinated beamforming.

\subsection{Related Work}

Stochastic geometry is a mathematical tool for characterizing the spatially averaged performance of wireless networks \cite{Baccelli2009,Haenggi2008}. By employing Poisson point processes (PPPs) to model the locations of BSs and users, this approach has provided valuable insights for the coverage and rate performances of various wireless network scenarios, including downlink cellular networks \cite{Guo2013,Andrews2011,Dhillon2012,Heath2013}, uplink cellular networks \cite{ElSawy2014,Di2016,Novlan2013}, relay-aided cellular networks \cite{Ganti2012,Elkotby2015}, device-to-device underlaid cellular networks \cite{ Lee2015}, multi-antenna systems \cite{Lee2015-2,Park2016,Tanbourgi2015}, mmWave networks\cite{Bai2015,Singh2015,Di2015}, and UAV networks\cite{Chetlur2017,Alkama2022}.

Recently, the analysis of satellite networks using stochastic geometry tools has gained significant momentum. For instance, the coverage probability of LEO satellite networks was provided by employing a binomial point process (BPP) to model the locations of satellites on the spherical surfaces \cite{Okati2020,Talgat2020,Talgat2021}. In \cite{Okati2020}, the coverage probability and rate were analytically derived by assuming a fixed number of satellites uniformly distributed on the surface of a sphere. The work in \cite{Talgat2020} characterized a contact distance distribution and nearest distributions considering multiple altitudes of satellites. Building on this, the downlink coverage probability was provided where satellites act as relays between users and LEO satellites in \cite{Talgat2021}.

In addition to the BPP-based modeling and analysis, the PPP-based performance analysis for satellite networks was proposed in \cite{Okati2022,Al2021,Park2023,Chae2023,park2023_2,lee2022}. In \cite{Okati2022}, LEO satellite networks' coverage and rate performance were investigated by modeling satellite locations according to a non-homogeneous PPP with a non-uniform satellite intensity according to the latitudes. The downlink coverage probability of satellite networks is provided by deriving the contact angle distribution in \cite{Al2021}. In \cite{Park2023}, the downlink coverage probability is provided by computing the conditional coverage probability according to the visible number of satellites. Moreover, the exact and approximated formulas for the coverage probability and the ergodic rate were derived as a function of the beamwidth by considering directional beamforming in \cite{Chae2023}. In \cite{park2023_2}, the rate coverage probability analysis for satellite-terrestrial integrated networks was proposed. Further, in \cite{lee2022}, the coverage probability was derived in terms of the orbit-geometry parameters.  

By modeling satellite networks with BPPs or PPPs, the prior work was able to offer insights regarding the coverage performance of densely deployed LEO satellite networks. Nonetheless, it is limited in that satellite coordination methods to mitigate the interference have not been counted. This is the main focus of our paper.

\subsection{Contributions}
In this paper, we propose a coverage analysis framework for the downlink satellite networks, wherein the dynamic coordinated beamforming is applied for interference mitigation. In particular, we model the locations of satellites and ground stations as independent homogeneous PPPs on the surface of the satellite sphere and the surface of the Earth, respectively. In our setup, we form a dynamic clustering set by including $K$ nearest satellites to the typical ground station. The set of ground stations within the same cluster is served through coordinated beamforming by which the intra-cluster interference is eliminated. The major contributions of this paper are summarized as follows.
\begin{itemize}
    \item To reflect the dynamic coordinated beamforming effect into the coverage analysis, we derive a distribution regarding the distance between the $K$th nearest satellite's location and the typical ground station's location, given the in-cluster relative distances. Unlike previous approaches that compute nearest distance distributions, we compute the $K$ nearest distance distribution since the interferences come from the satellites farther than the $K$th nearest satellite's location using dynamic coordinated beamforming strategy when at least $K$ satellites exist. The obtained distribution of $K$ nearest distance is a generalized distance distribution, encompassing the conditional distribution of the nearest satellite distance introduced in \cite{Park2023}, and it is also a generalization of the $K$ nearest distance distribution derived in a 2D plane \cite{Haenggi2005} to a spherical cap geometry. Further, different from the $K$ nearest distance distribution derived in \cite{Lee2015-2}, our approach considers the feasible region that the $K$th nearest satellite exists given the in-cluster relative distances since the visible spherical cap is a finite space.

\begin{figure}[t]
\begin{center}
\subfigure{\includegraphics[width=7.5cm]{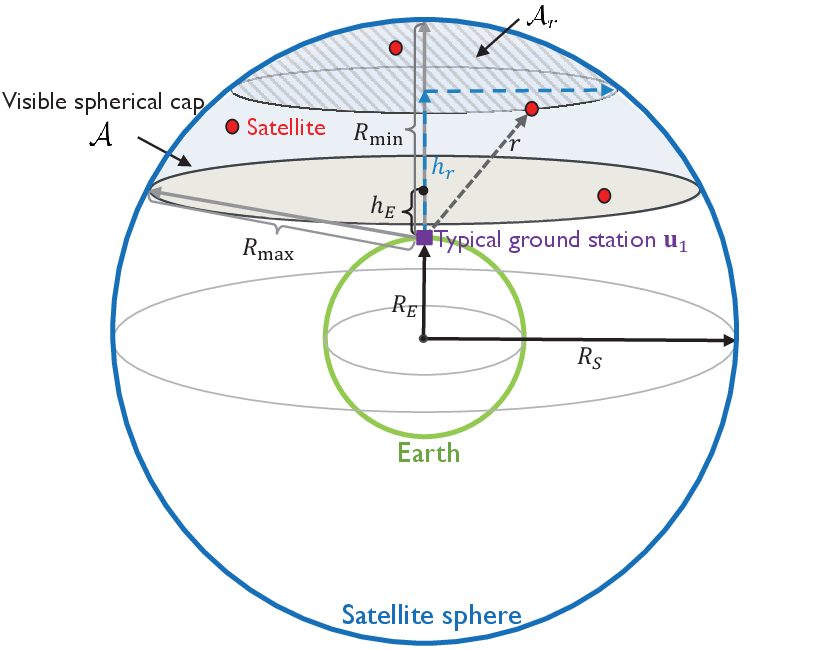}}
\end{center}
\caption{An illustration of satellite downlink networks, where the locations of the ground stations and satellites are modeled as independent PPPs.}\label{fig:network}
\end{figure}

    \item Leveraging the $K$ nearest distance distribution, we present an exact expression for the coverage probability conditioned on the relative distance. Our analysis includes a case that the number of visible satellites is less than $K-1$. In this case, the dynamic coordination set includes all the visible satellites, thus no interfering satellite exists. Given this situation, we obtain the signal-to-noise ratio (SNR) coverage instead of the signal-to-interference-plus-noise ratio (SINR) coverage. We derive the coverage probability conditioned on the visibility of at least $K$ satellites from the typical ground station, given the relative distance. Finally, we obtain relative distance-conditioned coverage probability by marginalizing the coverage probability for the number of visible satellites. This coverage probability captures the average behavior in all possible cluster geometries sharing a common relative distance. We further obtain an approximation of the relative distance-conditioned coverage probability for traceability. 

\begin{figure*}[t]
\begin{center}
\includegraphics[width=14.5cm]{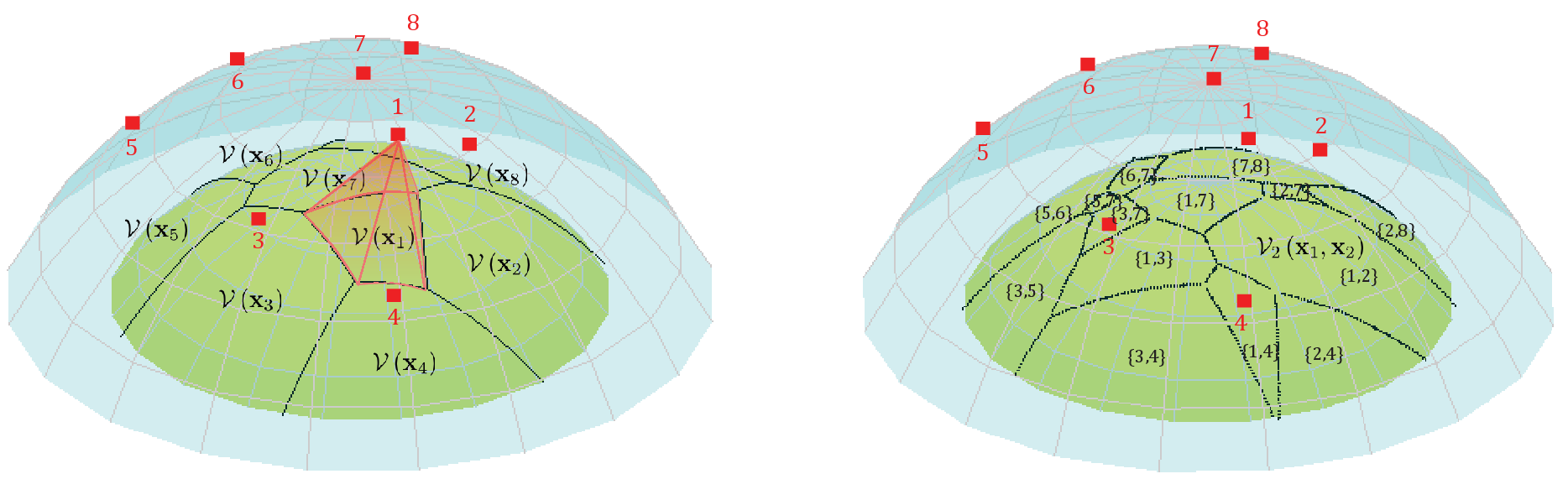}
\end{center}
\caption{The example of the first-order Voronoi tessellation on a three-dimensional plane for $N=8$ (left) and the second-order Voronoi tessellation with the same satellite locations (right).}
\label{fig:voronoi}
\end{figure*}
    
    \item Next, we provide the coverage probability marginalized over relative distance. To obtain this, we first derive the distribution of the relative distance. Then, the relative distance conditioned coverage probability is marginalized with respect to the relative distance distribution. This coverage probability provides an average behavior of coordinated beamforming.
    \item Through simulation, the accuracy of the derived coverage probabilities is verified. Our finding is that coordinated beamforming is more beneficial than without coordinated beamforming when the number of transmit antennas of the satellite is large, and the in-cluster relative distance is small. Further, we find that the optimal cluster size increases with satellite density if the number of transmit antennas is large enough.
    
\end{itemize}

\section{System Model}

This section explains the downlink satellite networks and dynamic satellite clustering model. Then, we describe the signal models and define the performance metrics for analyzing the downlink satellite networks considering the coordinated beamforming.

\subsection{Satellite Network Model}

\subsubsection{Spatial Distribution of Satellites and Ground Stations}
We consider downlink satellite networks, where each satellite is equipped with $N_t$ antennas, and each downlink ground station is equipped with $N_r$ antennas. The satellites are located on the surface of the sphere $\mathcal{S}_{R_{\sf S}}$ with radius $R_{\sf S}$ and the downlink ground stations are located on the surface of the Earth $\mathcal{S}_{R_{\sf E}}$ with radius $R_{\sf E}$. The locations of these satellites are established according to a homogeneous spherical Poisson point process (SPPP) with density $\lambda$, i.e., $\Phi=\{\mathbf{x}_1,\ldots,\mathbf{x}_N\}$, where $N$ follows Poisson random variable with mean $4\lambda \pi R_{\sf S}^2$. Further, the locations of downlink ground stations are assumed to be distributed according to a homogeneous SPPP, $\Phi_{\sf U}=\{\mathbf{u}_1,\ldots,\mathbf{u}_M\}$, which has density $\lambda_{\sf U}$. We assume that $\Phi_{\sf U}$ is independent of $\Phi$.

\subsubsection{Visible Spherical Cap}
By Slivnyak's theorem \cite{Haenggi2005}, we only consider the typical ground station located at $(0,0, R_{\sf E})$. Further, we assume that the satellites located above altitude $h_{\sf E}$ from the typical ground station are visible to the typical ground station. Then, as depicted in Fig. \ref{fig:network}, the visible area to the typical ground station is the portion of the sphere formed by the intersection of the tangent plane at the altitude $h_{\sf E}$ above the typical ground station and the satellites’ sphere. We define this visible area as a visible spherical cap $\mathcal{A}$, and the visible satellites on the visible spherical cap are distributed according to a PPP with density $\lambda |\mathcal{A}|$. The area of the visible spherical cap is computed by Archimedes' hat theorem \cite{cundy1989} as
\begin{align}
    |\mathcal{A}|=2\pi(R_{\sf S}-R_{\sf E}-h_{\sf E})R_{\sf S}.
\end{align}
Further, we define a spherical cap $\mathcal{A}_r$ that encompasses all points whose distance from the typical ground station's position is less than $r$ for $R_{\sf min} \le r \le R_{\sf max}$, where $R_{\sf min}=R_{\sf S}-R_{\sf E}$ and $R_{\sf max}=\sqrt{R_{\sf S}^2-R_{\sf E}^2-2R_{\sf E}h_{\sf E}}$. Then the area of the spherical cap $\mathcal{A}_r$ is given by
\begin{align}
    |\mathcal{A}_r| = 2\pi (R_{\sf S}-R_{\sf E}-h_r)R_{\sf S},
\end{align}
where $h_r=\frac{(R_{\sf S}^2-R_{\sf E}^2)-r^2}{2R_{\sf E}}$ is the closest distance from the typical ground station and the plane of the spherical cap $\mathcal{A}_r$.

\subsection{Dynamic Clustering Model}
We consider the dynamic clustering for coordinated beamforming \cite{Lee2015}, where the typical ground station coordinates with the $K$ nearest satellites. From the network point of view, the set of ground stations' locations coordinated by their $K$ nearest satellites is represented by the $K$th-order Voronoi tessellation. The $K$th-order Voronoi cell $\mathcal{V}_K(\mathbf{x}_1,\ldots,\mathbf{x}_K)$ is defined as the set of points on the surface of the Earth that have the same $K$ nearest satellites $\{\mathbf{x}_1,\ldots,\mathbf{x}_K\}$, i.e.,
\begin{align}
     \mathcal{V}_K(\mathbf{x}_1,\ldots,\mathbf{x}_K) \!=\! \cap_{n=1}^K\! \{\mathbf{u}\in \mathcal{S}_{R_{\sf E}} | \norm{\mathbf{u}-\mathbf{x}_n}_2 \!\le\! \norm{\mathbf{u}-\mathbf{x}_\ell}_2 \},
\end{align}
where $\mathbf{x}_\ell \in \Phi/\{\mathbf{x}_1,\ldots,\mathbf{x}_K\}$. As an example of $K=2$, the second-order Voronoi tessellation is illustrated in Fig. \ref{fig:voronoi}. The ground stations in $\mathcal{V}(\mathbf{x}_1,\mathbf{x}_2)$ are served by the cluster formed by satellites $\mathbf{x}_1$ and $\mathbf{x}_2$. We assume that each satellite serves one ground station per time-frequency resource, and the rest of the ground stations are served with different time-frequency resources. 

We further define the relative distance between the typical ground station and the nearest satellite, which is normalized by the distance to the $K$th nearest satellite as 
\begin{align}
    \delta = \frac{\norm{\mathbf{x}_1-\mathbf{u}_1}_2}{\norm{\mathbf{x}_K-\mathbf{u}_1}_2},
\end{align}
for $\frac{R_{\sf min}}{R_{\sf max}}\le \delta \le 1$. This relative distance plays a significant role in understanding the gain of coordinated beamforming according to the specific in-cluster geometry. The large $\delta$ indicates a large feasible region for the interfering satellites and vice versa. Therefore, when $\delta \simeq 1$, it implies that the ground station is in the inner part of the cell, whereas when $\delta \simeq \frac{R_{\sf min}}{R_{\sf max}}$, it suggests that the ground station is at the edges of the cell. 


\subsection{Coordinated Beamforming Gain}


For coordinated beamforming, we consider an adaptive beamforming technique that aims to steer the main beam of an antenna array toward a desired signal while simultaneously suppressing interference to other directions under a uniform linear array (ULA) antenna structure \cite{Li2003}. 
We assume that each ground station has knowledge of the angles of arrival (AoA) from the $K$ satellites within the cluster. 
Based on the AoA information, the coordinated beamforming constructs directional beamforming towards the desired direction while nullifying the $K-1$ intra-cluster interference signals when $N_t \ge K$. 

We now explain the model of the transmit and receive beamforming gains from satellites to the typical ground station. Specifically, we consider a two-lobe approximation of the antenna radiation pattern as in \cite{Bai2014, Renzo2015}. Let $G_n$ be the effective antenna gain from the $n$th satellite to the typical ground station. We assume the receive beam of the typical ground station is perfectly oriented to the transmit beam of the nearest satellite, while it is not aligned to the transmit beam of the interfering satellites. Then, the effective antenna gain from satellite $n$ in dB scale is modeled as
\begin{align}
    G_n = \begin{cases}
  G_{\sf tx}^{\sf (main)}+G_{\sf rx}^{\sf (main)}+20\log_{10}\left(\frac{c}{4\pi f_c}\right)  & n = 1 \\
  0 & 2\le n \le K \\
  G_{\sf tx}^{\sf (side)}+G_{\sf rx}^{\sf (side)}+20\log_{10}\left(\frac{c}{4\pi f_c}\right) & K+1\le n
\end{cases}, 
\end{align}
where $G_{\sf tx}^{\sf (main)}$ is the transmit antenna gain of the satellite for the main lobes, $G_{\sf rx}^{\sf (main)}$ is the receive antenna gain of the typical ground station for the main lobes, $G_{\sf tx}^{\sf (side)}$ and $G_{\sf rx}^{\sf (side)}$ are the transmit and receive antenna gain for the side lobes, respectively, $f_c$ is the carrier frequency, and $c$ is the speed of light.

Considering the coordinated beamforming, we model the main lobe beam gain of the transmit antenna as
\begin{align}
    G_{\sf tx}^{\sf (main)} = G_0+10\log_{10}(N_t-K+1),
\end{align}
where $G_0$ represents the main lobe beam gain for the single antenna. 
 
\subsection{Channel Model}
We adopt a time-honored Shadowed-Rician satellite channel model that can accurately model the compound effects of satellite channels' medium- and small-scale fading effects in various frequency bands \cite{Abdi2003}. We use $\sqrt{H_n}\in \mathbb{C}$ to represent the complex channel coefficient from the $n$th satellite to the typical ground station. The probability density function (PDF) of the $\sqrt{H_n}$ is given by
\begin{align}
    f_{\sqrt{H_n}}(x)&=\left(\frac{2bm}{2bm+\Omega}\right)^m \frac{x}{b}\exp\left(\frac{-x^2}{2b}\right) \nonumber\\ &~~~~~~~~~~~~\cdot F_1\left(m;1;\frac{\Omega x^2}{2b(2bm+\Omega)}\right),
\end{align}
where $F_1(a;b;c)$ is the confluent hyper-geometric function of the first kind \cite{magnus1967}, $2b$ and $\Omega$ are the average power of the scatter component and line-of-sight (LOS) component, respectively, and $m$ is the Nakagami parameter.

\subsection{Signal Model}
We assume that each ground station received an information symbol from the nearest satellite. Specifically, we consider that the $n$th satellite sends an information symbol $s_n$ to the intended ground station. Let $\mathbf{u}_1$ be the location of the typical ground station located at $(0,0,R_{\sf E})$. Then, we define the path-loss model from the $n$th satellite to the typical ground station as $\norm{\mathbf{x}_n-\mathbf{u}_1}^{-\frac{\alpha}{2}}$, where $\alpha$ is a path-loss exponent. 

Using the defined network and channel models, the received signal at the typical ground station can be expressed as
\begin{align}
    &y_1 = \sqrt{G_1}\norm{\mathbf{x}_1-\mathbf{u}_1}^{-\frac{\alpha}{2}}\sqrt{H_1} s_1 \nonumber\\ &+ \sum_{n=2}^{K} \sqrt{G_n}\norm{\mathbf{x}_n-\mathbf{u}_1}^{-\frac{\alpha}{2}}\sqrt{H_n} s_n \nonumber \\
    & +\!\! \sum_{\mathbf{x}_n \in \Phi \cap \mathcal{A} / \{\mathbf{x}_1,\ldots,\mathbf{x}_K\}} \!\!\sqrt{G_n}\norm{\mathbf{x}_n-\mathbf{u}_1}^{-\frac{\alpha}{2}}\sqrt{H_n}s_n + z_1,
\end{align}
where $z_1$ denotes the additive Gaussian noise, i.e., $z_1\sim \mathcal{CN}(0,\sigma^2)$ and the transmit symbol $s_n$ is assumed to be drawn from $\mathcal{CN}(0,P)$ to meet the average power constraint $\mathbb{E}\left[|s_n|^2\right]=P$.
Then, the SINR for the typical ground station is given by

\begin{align}
    {\sf SINR} &= \frac{G_1 P H_1 \norm{\mathbf{x}_1-\mathbf{u}_1}^{-\alpha}}{\sum_{\mathbf{x}_n \in \Phi \cap \mathcal{A} / \{\mathbf{x}_1,\ldots,\mathbf{x}_K\}}G_n P H_n \norm{\mathbf{x}_n-\mathbf{u}_1}^{-\alpha} + \sigma^2} \nonumber \\
    &= \frac{ H_1 (\delta\norm{\mathbf{d}_K})^{-\alpha}}{\sum_{\mathbf{x}_n \in \Phi \cap \mathcal{A} / \{\mathbf{x}_1,\ldots,\mathbf{x}_K\}} \Bar{G}_n H_n\norm{\mathbf{d}_n}^{-\alpha} +\Bar{\sigma}^2 }\nonumber\\ & =\frac{ H_1 (\delta\norm{\mathbf{d}_K})^{-\alpha}}{I_K +\Bar{\sigma}^2 },
\end{align}
where $\norm{\mathbf{x}_n-\mathbf{u}_1} = \norm{\mathbf{d}_n}$ is the distance from the $n$th nearest satellite to the typical ground station, $\delta = \frac{\norm{\mathbf{d}_1}}{\norm{\mathbf{d}_K}}$, $\Bar{G}_n = \frac{G_n}{G_1}$ and $\Bar{\sigma}^2 = \frac{\sigma^2}{P G_1}$ are the normalized antenna gains and noise power respectively. We assume that $\Bar{G}_n$ has the same value for $n\ne\{1,2,\ldots,K\}$. Further, $I_K$ is the normalized aggregated out-of-cluster interference power which is given by
\begin{align}
    I_K = \sum_{\mathbf{x}_n \in \Phi \cap \mathcal{A} / \{\mathbf{x}_1,\ldots,\mathbf{x}_K\}} \Bar{G}_n H_n\norm{\mathbf{d}_n}^{-\alpha}.
\end{align}

\subsection{Performance Metric}
We characterize the coverage probability at two different levels. 
\subsubsection{Specific In-Cluster Geometry}
We first characterize the average of the coverage probability over all possible satellites' geometries with a given relative distance $\delta$ as 
\begin{align}
    &P_{{\sf SINR}|\delta}^{\sf Cov}(K, N_t, \alpha,\Tilde{\lambda}, R_{\sf min}, R_{\sf max}, \delta  ; \gamma)    = \mathbb{P}\left[{\sf SINR}\ge \gamma |\delta\right] \nonumber \\
    &=\mathbb{P}\left[\frac{ H_1 (\delta\norm{\mathbf{d}_K})^{-\alpha}}{\sum_{\mathbf{x}_n \in \Phi \cap \mathcal{A} / \{\mathbf{x}_1,\ldots,\mathbf{x}_K\}} \Bar{G}_n H_n\norm{\mathbf{d}_n}^{-\alpha} +\Bar{\sigma}^2 } \ge \gamma \bigg| \delta \right]. \label{eq:delta_cond_cov}
\end{align} 
This conditional coverage probability represents the average behavior of all possible in-cluster geometries that share a specific value of $\delta$. 

\subsubsection{Average In-Cluster Geometry}
Further, we compute the average of the coverage probability over all possible satellites' geometries by marginalizing over $\delta$ as
\begin{align}
    &P_{{\sf SINR}}^{\sf Cov}(K, N_t, \alpha,\Tilde{\lambda}, R_{\sf min}, R_{\sf max}  ; \gamma)    = \mathbb{P}\left[{\sf SINR}\ge \gamma \right] \nonumber \\
    &=\mathbb{E}_{\delta}\!\left[\!\mathbb{P}\!\left[\!\frac{ H_1 (\delta\norm{\mathbf{d}_K})^{-\alpha}}{\sum_{\mathbf{x}_n \in \Phi \cap \mathcal{A} / \{\mathbf{x}_1,\ldots,\mathbf{x}_K\}} \Bar{G}_n H_n\norm{\mathbf{d}_n}^{-\alpha} \!+\!\Bar{\sigma}^2 } \!\ge\! \gamma \bigg| \delta \right]\!\right]. \label{eq:delta_avg_cov}
\end{align}
This coverage probability represents the average behavior of coordinated beamforming. 

\section{Coverage Probability Analysis for Specific Relative In-Cluster Geometry}
In this section, we provide the coverage probability in \eqref{eq:delta_cond_cov} and then derive an upper and lower bound of the coverage probability.  

\begin{figure}[t]
    \centering
    \subfigure{\includegraphics[width=8.6cm]{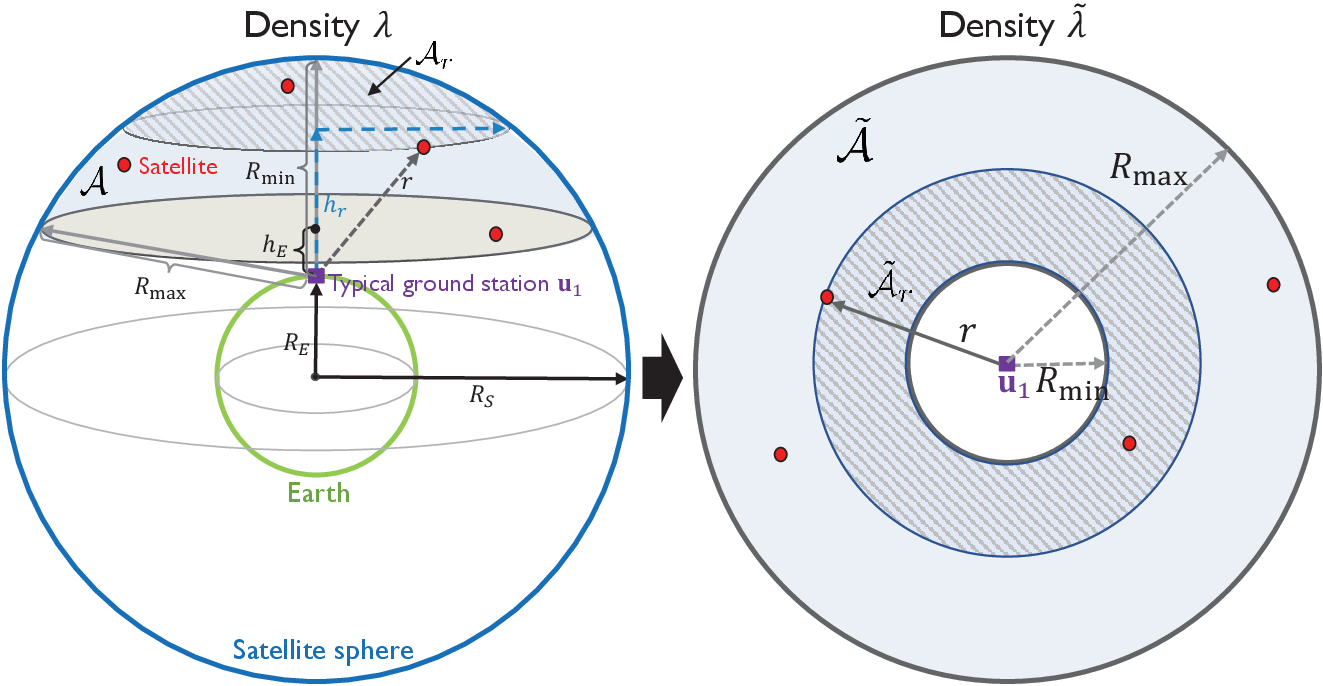}}
    \caption{The satellite network transformed to $\mathbb{R}^2$.} \label{fig:2Dnet}
\end{figure}

\subsection{A Replacement Lemma}
We introduce the replacement of satellite networks' geometric representation from $\mathbb{R}^3$ sphere to $\mathbb{R}^2$ planes. Specifically, we define the replaced $\mathbb{R}^2$ plane denoted as $\tilde{\mathcal{A}}$, where $\tilde{\mathcal{A}}$ is a circular ring with outer and inner radii are $R_{\sf min}$ and $R_{\sf max}$, which is illustrated Fig. \ref{fig:2Dnet}. In this area, the locations of satellites are established according to the homogeneous PPP with density $\Tilde{\lambda}$. 
\begin{lemma}[A Replacement Lemma]
    By letting $\Tilde{\lambda}=\lambda \frac{R_S}{R_E}$, the point process of satellites' locations in the visible spherical cap $\mathcal{A}$ is identical to the point process with points' location within circular ring $\tilde{\mathcal{A}}$.
    \begin{proof}
        See Appendix A.
    \end{proof}
\end{lemma}

The introduced replacement lemma allows us to easily analyze the coverage performance of satellite networks in a 2D circular ring plane instead of a 3D sphere. Using this, we consider the typical ground station located at the center of the ring and use the converted representation of PPP parameters as
$\Tilde{\lambda}= \lambda \frac{R_S}{R_E}$, $|\mathcal{\tilde{A}}| = \pi(R_{\sf max}^2 - R_{\sf min}^2)$, and $|\mathcal{\tilde{A}}_r| = \pi(r^2- R_{\sf min}^2)$. This representation will be used in the sequel.

\subsection{Statistical Properties}
We present a set of significant lemmas to derive the general characterization of coverage probability.

\begin{lemma}[The conditional nearest satellite distance distribution]
    For $K\ge 2$, the conditional distribution of the distance $\norm{\mathbf{d}_1}$ between a typical ground station and the nearest satellite is given by
    \begin{align}
         &f_{\norm{\mathbf{d}_1}|1\le\Phi(\mathcal{A})\le K-1}(r_1) \nonumber\\&= \frac{\sum_{u=1}^{K-1}(\Tilde{\lambda}\pi )^u\left(R_{\sf max}^2-r_1^2\right)^{u-1}2r_1/(u-1)!}{\sum_{n=1}^{K-1}( \Tilde{\lambda}\pi)^n\left(R_{\sf max}^2-R_{\sf min}^2\right)^n/n!}.
    \end{align}
    \begin{proof}
        See Appendix B.
    \end{proof}


\end{lemma}

\begin{lemma}[The conditional $K$th nearest satellite distance distribution]
    For $K\ge 2$, the conditional distribution of the distance $\norm{\mathbf{d}_K}$ between a typical ground station and the $K$th nearest satellite is given by
    \begin{align}
    &f_{\norm{\mathbf{d}_K} | \Phi(\mathcal{A}) \ge K, \delta} (r_K)= \frac{1}{v(\Tilde{\lambda},R_{\sf min},R_{\sf max},\delta,K)}\nonumber\\&\cdot \sum_{i=0}^{K-1}\Bigg[ \frac{2r_K i(\Tilde{\lambda}\pi)^i (r_K^2 - R_{\sf min}^2)^{i-1}}{i!}\Bigg(e^{-\Tilde{\lambda}\pi(R_{\sf max}^2-R_{\sf min}^2)}\nonumber\\ &\cdot\sum_{t=0}^{K-i-1}\frac{(\Tilde{\lambda}\pi)^t(R_{\sf max}^2-r_K^2)^t}{t!}-e^{-\Tilde{\lambda}\pi(r_K^2-R_{\sf min}^2)}\Bigg) \nonumber\\ &+\frac{(\Tilde{\lambda}\pi)^i (r_K^2 - R_{\sf min}^2)^{i}}{i!} \Bigg(2\pi\Tilde{\lambda}r_K e^{-\Tilde{\lambda}\pi(r_K^2-R_{\sf min}^2)}\nonumber\\ &-2r_Ke^{-\Tilde{\lambda}\pi(R_{\sf max}^2-R_{\sf min}^2)}\sum_{t=0}^{K-i-1}\frac{t(\Tilde{\lambda}\pi)^t(R_{\sf max}^2-r_K^2)^{t-1}}{t!}\Bigg) \Bigg], \label{eq:K_nearest_dist}
\end{align}
where
\begin{align}
    &v(\Tilde{\lambda},R_{\sf min},R_{\sf max},\delta,K)\nonumber\\ &=\sum_{j=0}^{K-1}\frac{(\Tilde{\lambda}\pi)^j(\frac{R_{\sf min}^2}{\delta^2}-R_{\sf min}^2)^j}{j!} \Bigg(e^{-\Tilde{\lambda}\pi(\frac{R_{\sf min}^2}{\delta^2}-R_{\sf min}^2)}\nonumber\\ &-e^{-\Tilde{\lambda}\pi(R_{\sf max}^2-R_{\sf min}^2)}\sum_{w=0}^{K-j-1}\frac{(\Tilde{\lambda}\pi)^w(R_{\sf max}^2-\frac{R_{\sf min}^2}{\delta^2})^w}{w!}\Bigg),
\end{align}
for $\frac{R_{\sf min}}{\delta}\le r_K \le R_{\sf max}$.
    \begin{proof}
        See Appendix C.
    \end{proof}

\label{lem:Knearest}
\end{lemma}

The resulting conditional distribution is the generalized distance distribution. Especially when $K=1$ and $\delta=1$, we obtain the conditional distribution of the nearest satellite distance as
\begin{align}
    f_{\norm{\mathbf{d}_1} | \Phi(\mathcal{A}) \ge 1} (r_1)=\frac{2\Tilde{\lambda}\pi e^{\Tilde{\lambda}\pi R_{\sf min}^2}}{1-e^{-\Tilde{\lambda}\pi(R_{\sf max}^2-R_{\sf min}^2)}}r_1 e^{-\Tilde{\lambda}\pi r_1^2},
\end{align}
which is the same as obtained in \cite{Park2023}. It's important to note that the derived conditional distribution depends on the $\delta$, while the $K$th nearest distribution obtained in \cite{Lee2015-2} is invariant with respect to $\delta$. This discrepancy arises because the visible spherical cap is a finite area. Conditioning on the value of $\delta$ determines the feasible region where the $K$th nearest satellite exists.

\begin{lemma}[Laplace transform of the aggregated interference power plus noise power]
    For $K\ge 2$, we define the aggregated interference power from the out-of-cluster satellites when the $\norm{\mathbf{d}_K}=r_K$ as
    \begin{align}
        I_{r_K}=\sum_{\mathbf{x}_n \in \Phi \cap \mathcal{A}/\mathcal{A}_{r_K}}\Bar{G}_n H_n \norm{\mathbf{d}_n}^{-\alpha}.
    \end{align}
    Then, the conditional Laplace transform of aggregated interference power plus noise power is given by
    \begin{align}
    &\mathcal{L}_{I_{r_K}+\Bar{\sigma}^2|\Phi(\mathcal{A})\ge K}(s) 
    =  \exp \Bigg(-s\Bar{\sigma}^2  \nonumber\\& -2 \pi \tilde{\lambda}  \int_{r_K}^{R_{\sf max}}\Bigg[1-\sum_{z=0}^{m-1} \zeta(z) \frac{\Gamma(z+1)}{(\beta-c+s \Bar{G}_n v^{-\alpha})^{z+1}}\Bigg]v\mathrm{d}v \Bigg), \label{eq:Laplace}
    \end{align}
    where $\beta = \frac{1}{2b}$, $c=\frac{\Omega}{2b(2bm+\Omega)}$, $\zeta(z)=\left(\frac{2bm}{2bm+\Omega}\right)^m \frac{\beta (-1)^z (1-m)_z c^z}{(z!)^2}$, $(x)_z=x(x+1)\cdots(x+z-1)$ is the Pochhammer symbol \cite{gradshteyn2014}, and $\Gamma(z)=(z-1)!$ is the Gamma function for positive integer $z$.
    \begin{proof}
        See Appendix D.
    \end{proof}
\end{lemma}


\subsection{Exact Expression}
The following theorem provides an exact expression for conditional SINR coverage probability.
\begin{theorem}
    For $K\ge 2$, the conditional SINR coverage probability for a given $\delta$ is obtained by
    \begin{align}
        &P_{{\sf SINR}|\delta}^{\sf Cov}(K, N_t, \alpha,\Tilde{\lambda}, R_{\sf min}, R_{\sf max}, \delta  ; \gamma)\nonumber \\ &=\sum_{n=1}^{K-1}\frac{(\Tilde{\lambda}\pi )^n\left(R_{\sf max}^2-R_{\sf min}^2\right)^n}{n!} e^{-\Tilde{\lambda}\pi\left(R_{\sf max}^2-R_{\sf min}^2\right)} \nonumber\\ &\cdot \! \int_{R_{\sf min}}^{R_{\sf max}} \Bigg[1-\sum_{z=0}^{m-1}\frac{\zeta(z)\Gamma(z)}{(\beta-c)^{z+1}}\Bigg(1-
        e^{-(\beta-c)\gamma r_1^{\alpha}\Bar{\sigma}^2}\nonumber\\&~~~~~~~~ \cdot\sum_{v=0}^{z}\frac{((\beta-c)\gamma r_1^\alpha \Bar{\sigma}^2)^v}{v!}\Bigg) \Bigg] f_{\norm{\mathbf{d}_1}|1 \! \le \! \Phi(\mathcal{A})\le K-1}(r_1)\mathrm{d}r_1 \nonumber\\ &+ \left(1-\sum_{n=0}^{K-1}\frac{(\Tilde{\lambda}\pi )^n\left(R_{\sf max}^2-R_{\sf min}^2\right)^n}{n!} e^{-\Tilde{\lambda}\pi\left(R_{\sf max}^2-R_{\sf min}^2\right)}\right) \nonumber \\ & \cdot\int_{\frac{R_{\sf min}}{\delta}}^{R_{\sf max}} \Bigg[1\!\!-\!\!\sum_{z=0}^{m-1}\frac{\zeta(z)\Gamma(z)}{(\beta\!-\!c)^{z+1}}\Bigg(1 \!-\!\sum_{v=0}^{z} \frac{((\beta\!-\!c)\gamma(\delta r_K)^\alpha)^v}{v!} \nonumber\\&~~~~~~~~~~~~~\cdot(-1)^v\frac{\mathrm{d}^v \mathcal{L}_{I_{r_K}+\Bar{\sigma}^2|\Phi(\mathcal{A})\ge K}(s)}{\mathrm{d}s^v} \bigg|_{s=(\beta-c)(\delta r_K)^{\alpha}\gamma} \Bigg)\Bigg] \nonumber\\ &~~~~~~~~~~~~~~~~~~~~~~~~~~~~~~\cdot f_{\norm{\mathbf{d}_K} | \Phi(\mathcal{A}) \ge K, \delta} (r_K)  \mathrm{d}r_K. \label{eq:theorem1}
    \end{align}
    \begin{proof}
        See Appendix E.
    \end{proof}
\end{theorem}

\begin{figure}[t]
    \centering
    \subfigure{\includegraphics[width=8cm]{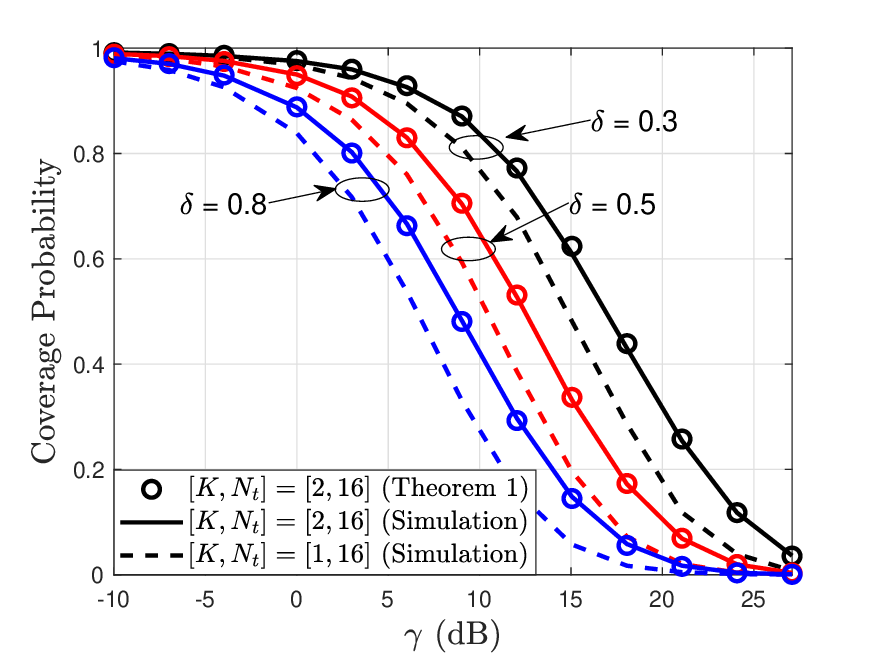}}
    \subfigure{\includegraphics[width=8cm]{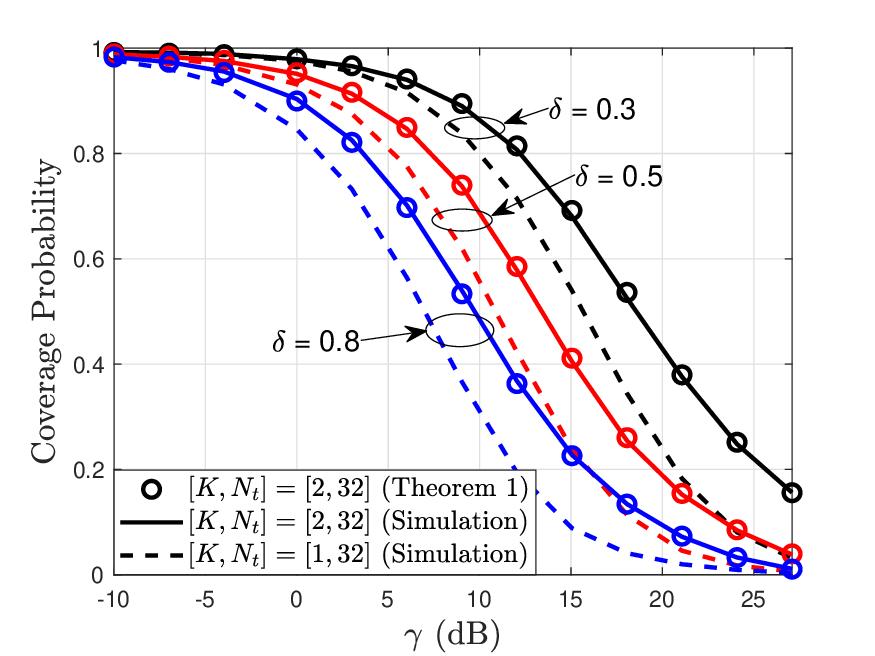}}
    \caption{The conditional coverage probability for $[m, b, \Omega]=[1,0.063,8.97 \times 10^{-4}]$ and $\lambda|\mathcal{A}|=5$.} \label{fig:delta_cond_A5}
\end{figure}

Fig. \ref{fig:delta_cond_A5} confirms that the derived conditional coverage probability in Theorem 1 exactly matches the numerical conditional coverage probability. We set $\alpha=2$,  $P=43$ dBm, $N_0=-174$ dBm/Hz, $W=100$ MHz, $G_0=20$ dBi, $f_c=13.5$ GHz, $\Bar{G}_n=0.1$, $R_{\sf S}-R_{\sf E}=500$ km, $h_{\sf E}=0$ km and $N_t=N_r$. These parameters remain consistent unless mentioned otherwise. When not using coordination ($K=1$), $\delta$ is defined as the nearest distance normalized by the second nearest distance.
As can be seen in the figures, the coordinated beamforming gain is large when the number of transmit antennas of the satellite is large. Further, the coordinated beamforming gain is large when $\delta$ is small since small $\delta$ implies that the area where out-of-cluster interfering satellites exist is small. From the network point of view, coordinated beamforming provides more advantages for the ground stations in the inner part of the clusters, and conversely, less advantages for those near the cell edges. This is because the ground stations in the near edges of cells still suffer from substantial interference even after mitigating in-cluster interference, due to the strong out-of-cluster interference.

\subsection{Approximation}
Although Theorem 1 provides general characterization, the derivative of the Laplace transform hinders the compact expression. For more compact characterizations, we next provide an approximation of the conditional coverage probability with full generality. 

\begin{theorem}
    For $K\ge 2$, the conditional coverage probability given $\delta$ is approximated as
    \begin{align}
    & P_{{\sf SINR}|\Phi(\mathcal{A})\ge K,\delta}^{\sf Cov}(K, N_t, \alpha,\Tilde{\lambda}, R_{\sf min}, R_{\sf max},\delta; \gamma) \nonumber \\
    &\simeq 1-\sum_{z=0}^{m-1}\frac{\zeta(z)}{(\beta-c)^{z+1}}z!\sum_{\ell = 0}^{z+1} \binom{z+1}{\ell}(-1)^{\ell} \nonumber\\&~~~\cdot \int_{\frac{R_{\sf min}}{\delta}}^{R_{\sf max}}\mathcal{L}_{I_{r_K}+\Bar{\sigma}^2|\Phi(\mathcal{A})\ge K}\left(\ell  \kappa (\beta\!-\!c)\gamma (\delta r_K)^{\alpha}\right)\nonumber\\  &~~~~~~~~~~~~~~~~~~~~~~~~~~~~~\cdot f_{\norm{\mathbf{d}_K} | \Phi(\mathcal{A}) \ge K, \delta} (r_K)  \mathrm{d}r_K. \label{eq:approx}
\end{align} 
    where $ \left((z+1)!\right)^{-\frac{1}{z+1}}\le \kappa \le 1$.
    \begin{proof}
        See Appendix F.
    \end{proof}
\end{theorem}
The derived derived in Theorem 2 coincides with the exact coverage probability presented in Theorem 1 when $m=1$. Further for $m \ge 2$, we can obtain the tightly matched approximation for the coverage probability through a careful adjustment of the parameter $\kappa$.

\begin{figure}[t]
    \centering
    \subfigure[]{\includegraphics[width=8.1cm]{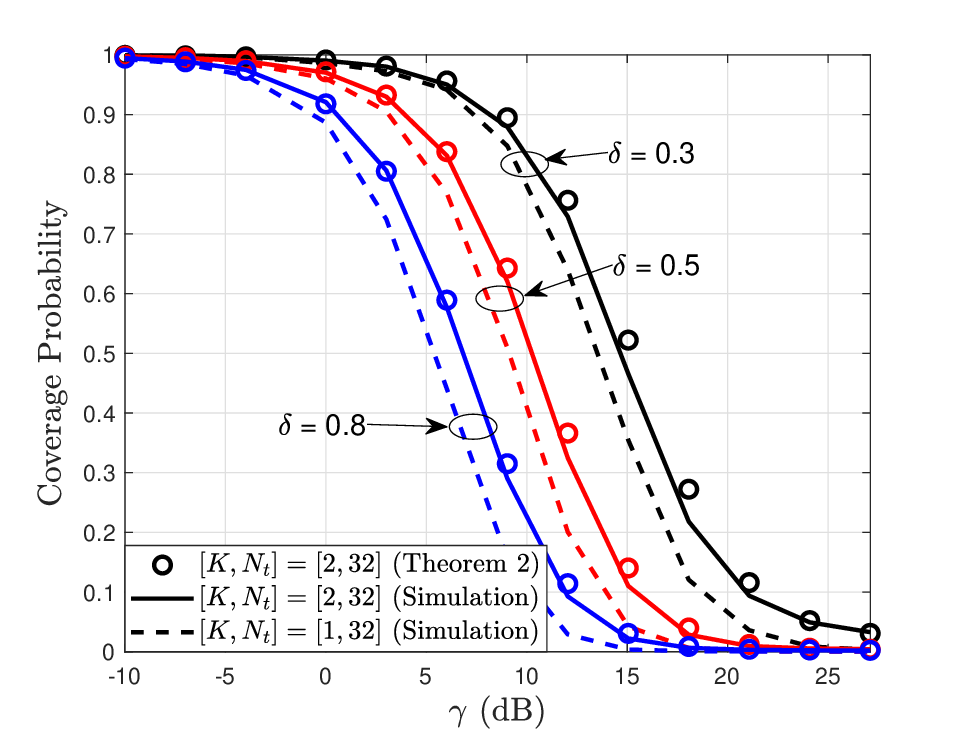}}
    \subfigure[]{\includegraphics[width=8.1cm]{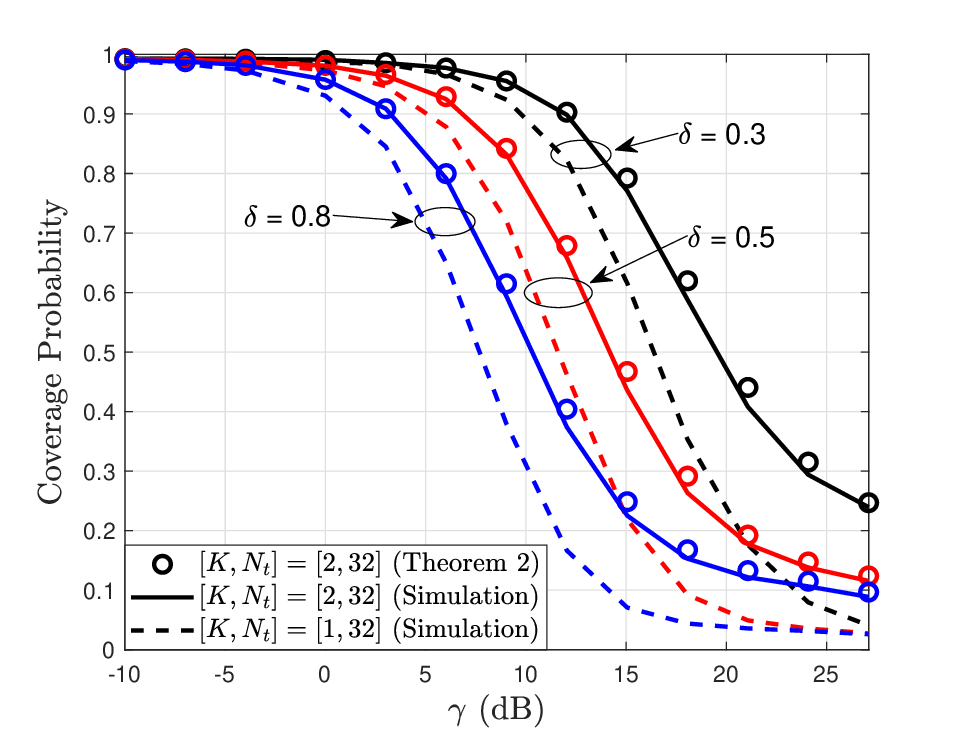}}
    \caption{The approximation of the conditional coverage probability for $[m, b, \Omega]=[10,0.126,0.835]$, (a) $\lambda|\mathcal{A}|=10$, and (b) $\lambda|\mathcal{A}|=5$. } \label{fig:upperbound}
\end{figure}

Fig. \ref{fig:upperbound} confirms that the approximation derived in \eqref{eq:approx} tightly matches the exact conditional coverage probability for the entire range of target SINRs, various $\delta$, and satellite densities. In this simulation, we set $\kappa = \left((z+1)!\right)^{-\frac{1}{z+1}}$.

\section{Coverage Probability Analysis \\for Average In-Cluster Geometry}
This section provides the coverage probability in \eqref{eq:delta_avg_cov}. Further, we characterize the optimal cluster size in terms of ergodic spectral efficiency. 


\subsection{General Characterization}
 To obtain a coverage probability averaged over the distribution of $\delta$, we first compute the probability density function (PDF) of $\delta$ using the following Lemma.
 
\begin{lemma}
For $K\ge 2$, the PDF of $\delta = \frac{\norm{\mathbf{d}_1}}{\norm{\mathbf{d}_K}}$ is given by
\begin{align}
    &f_{\delta|\Phi(\mathcal{A})\ge K}(x)\nonumber\\&=\frac{2e^{\Tilde{\lambda}\pi R_{\sf min}^2}x(1\!-\!x^2)^{K\!-\!2}\left[\Gamma\!\left(K,{\Tilde{\lambda}\pi \frac{R_{\sf min}^2}{x^2}}\right)\!-\!\Gamma\!\left(K,{\Tilde{\lambda}\pi R_{\sf max}^2}\right)\right]}{(K\!-\!2)! \left(1\!-\!\sum_{n=0}^{K-1}\frac{(\Tilde{\lambda}\pi )^n\left(R_{\sf max}^2-R_{\sf min}^2\right)^n}{n!} e^{-\Tilde{\lambda}\pi\left(R_{\sf max}^2-R_{\sf min}^2\right)}\right)}, \label{eq:delta_dist_lemma}
\end{align}
for $\frac{R_{\sf min}}{R_{\sf max}}\le x \le 1$, where $\Gamma(a,x)=\int_x^{\infty}t^{a-1}e^{-t}\mathrm{d}t$ denotes the incomplete gamma function.

\begin{proof}
    See Appendix G.
\end{proof} \label{lem:PDFdelta}
\end{lemma}


By applying the lemma \ref{lem:PDFdelta}, we obtain the SINR coverage probability averaged over the distribution of $\delta$ for $K \ge 2$ as the following theorem.

\begin{theorem}
For $K\ge2$, the SINR coverage probability is given by
      \begin{align}
        &P_{{\sf SINR}}^{\sf Cov}(K, N_t, \alpha,\Tilde{\lambda}, R_{\sf min}, R_{\sf max}  ; \gamma)\nonumber \\ &=\sum_{n=1}^{K-1}\frac{(\Tilde{\lambda}\pi )^n\left(R_{\sf max}^2-R_{\sf min}^2\right)^n}{n!} e^{-\Tilde{\lambda}\pi\left(R_{\sf max}^2-R_{\sf min}^2\right)} \nonumber\\ &\cdot \! \int_{R_{\sf min}}^{R_{\sf max}} \Bigg[1-\sum_{z=0}^{m-1}\frac{\zeta(z)\Gamma(z)}{(\beta-c)^{z+1}}\Bigg(1-
        e^{-(\beta-c)\gamma r_1^{\alpha}\Bar{\sigma}^2}\nonumber\\&~~~~~~~~ \cdot\sum_{v=0}^{z}\frac{((\beta-c)\gamma r_1^\alpha \Bar{\sigma}^2)^v}{v!}\Bigg) \Bigg] f_{\norm{\mathbf{d}_1}|1 \! \le \! \Phi(\mathcal{A})\le K-1}(r_1)\mathrm{d}r_1 \nonumber\\ &+ \left(1-\sum_{n=0}^{K-1}\frac{(\Tilde{\lambda}\pi )^n\left(R_{\sf max}^2-R_{\sf min}^2\right)^n}{n!} e^{-\Tilde{\lambda}\pi\left(R_{\sf max}^2-R_{\sf min}^2\right)}\right) \nonumber \\ & \cdot \int_{\frac{R_{\sf min}}{R_{\sf max}}}^{1}\! \int_{\frac{R_{\sf min}}{x}}^{R_{\sf max}} \!\Bigg[1\!\!-\!\!\sum_{z=0}^{m-1}\frac{\zeta(z)\Gamma(z)}{(\beta\!-\!c)^{z+1}}\Bigg(\!\!1 \!-\!\sum_{v=0}^{z} \frac{((\beta\!-\!c)\gamma(x r_K)^\alpha)^v}{v!} \nonumber\\&~~~~~~~~~~~~\cdot(-1)^v\frac{\mathrm{d}^v \mathcal{L}_{I_{r_K}+\Bar{\sigma}^2|\Phi(\mathcal{A})\ge K}(s)}{\mathrm{d}s^v} \bigg|_{s=(\beta-c)(x r_K)^{\alpha}\gamma} \Bigg)\Bigg] \nonumber\\ &~~~\cdot f_{\norm{\mathbf{d}_K} | \Phi(\mathcal{A}) \ge K,x} (r_K)  f_{\delta|\Phi(\mathcal{A})\ge K}(x)\mathrm{d}r_K  \mathrm{d}x. \label{eq:theorem1}
    \end{align}
    \begin{proof}
        See Appendix H.
    \end{proof}
\end{theorem}

Fig. \ref{fig:avgCov} shows that the derived SINR and rate coverage probabilities averaged over the distribution $\delta$ exactly match with numerical coverage probabilities. As shown in Fig. \ref{fig:avgCov}, the coordinated beamforming gain is larger than without satellite coordination when the satellite density is low for $K=2$. This is because the coordinated beamforming gain is relatively insignificant since a substantial number of interfering satellites still exist even after nullifying the one interfering signal when the satellite density is high.

\subsection{Optimal Cluster Size}

By assuming perfect channel state information (CSI) at the typical ground station, the ergodic spectral efficiency averaged over relative distance $\delta$ is computed as
\begin{align}
    &C(K, N_t, \alpha,\Tilde{\lambda}, R_{\sf min}, R_{\sf max}) \nonumber\\ &= \int_{0}^{\infty} \frac{\log_2 e}{1+\gamma}P_{{\sf SINR}}^{\sf Cov}(K, N_t, \alpha,\Tilde{\lambda}, R_{\sf min}, R_{\sf max}  ; \gamma)\mathrm{d}\gamma.
\end{align}
Then, the optimal cluster size $K^\star$ that maximizes the ergodic spectral efficiency is obtained by
\begin{align}
    K^\star =  \underset{K\in \{1,\ldots,N_t\}}{\max} C(K, N_t, \alpha,\Tilde{\lambda}, R_{\sf min}, R_{\sf max}).
\end{align}

\begin{figure}[t]
    \centering
    \subfigure[]{\includegraphics[width=8.1cm]{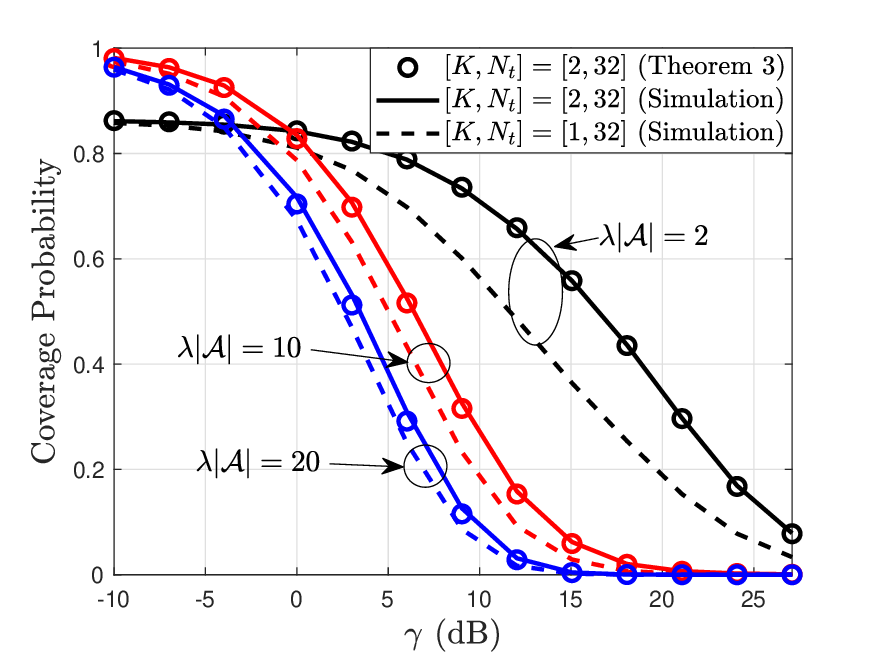}}
    \subfigure[]{\includegraphics[width=8.1cm]{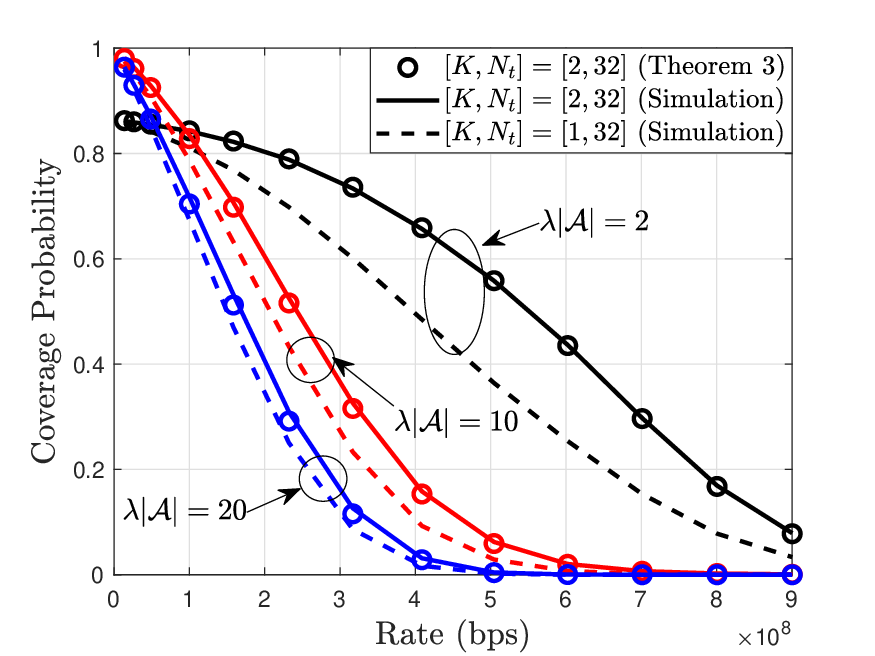}}
    \caption{The (a) SINR and (b) rate coverage probabilities for $[m, b, \Omega]=[1,0.063,8.97 \times 10^{-4}]$.} \label{fig:avgCov}
\end{figure}

\begin{figure}[t]
    \centering
    \subfigure{\includegraphics[width=8.1cm]{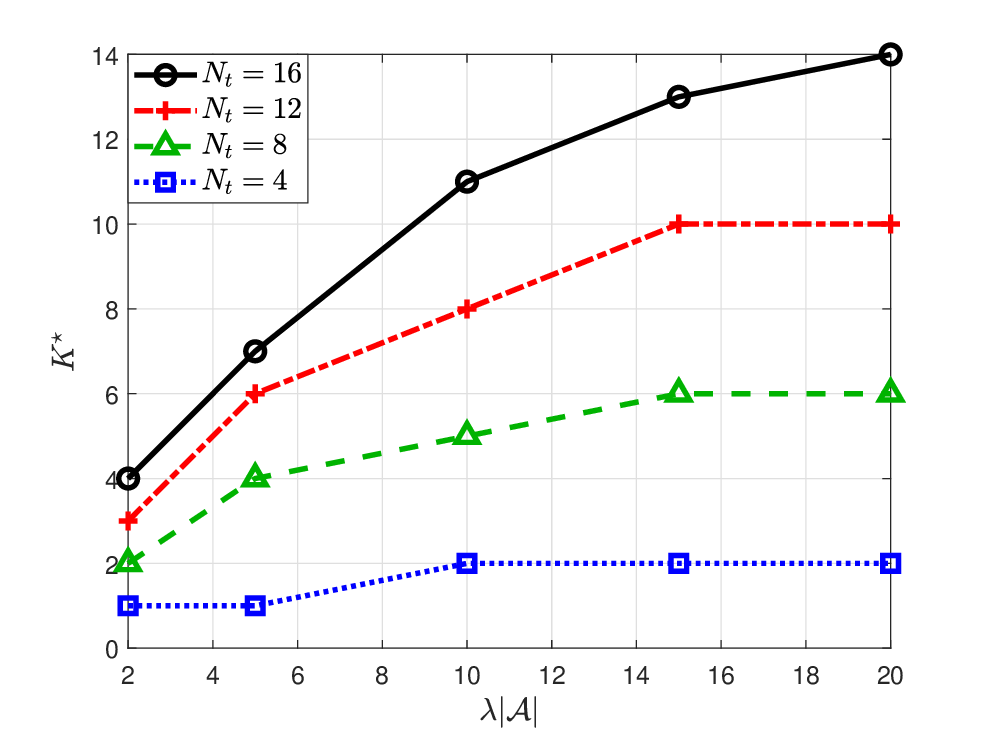}}
    \caption{The optimal cluster size $K^\star$ as a function of satellite density $\lambda|\mathcal{A}|$ for $[m, b, \Omega]=[1,0.063,8.97 \times 10^{-4}]$.} \label{fig:optK}
\end{figure}

Fig. \ref{fig:optK} illustrates optimal cluster size $K^\star$ as a function of satellite density $\lambda|\mathcal{A}|$. As can be seen, the optimal cluster size increases with satellite density when $N_t \ge 8$. This is because the interference becomes dominant in the dense network, necessitating more coordination. On the other hand, the optimal cluster size is between 2 and 5 in terrestrial networks using BS coordination in \cite{Lee2015-2}. The difference arises from the consideration of pilot overhead required for acquiring CSI in terrestrial networks,  whereas in downlink satellite networks between satellites and ground stations, we do not consider AoA acquisition overhead.
However, the optimal cluster size $K$ does not necessarily increase with the satellite density when $N_t$ is small. This phenomenon arises because canceling the $K-1$ interference costs $K-1$ degrees of freedom, leaving only $N_t-K$ degrees of freedom available for maximizing the signal power. The relative cost of interference cancellation is more pronounced when $N_t$ is small compared to when it is large. For instance, the optimal cluster size is $K^\star=1$ when $N_t=4$ and $\lambda|\mathcal{A}|=2,5$. In this case, it is better to use transmit antennas to maximize the desired signal power instead of canceling the interference signals. On the contrary, when $N_t$ is large, the optimal cluster size increases since the cost of interference cancellation is relatively negligible.

\section{Conclusion}
In this paper, we have investigated the coverage performances of dynamic coordinated beamforming for satellite downlink networks. Using stochastic geometry tools, we first derived the coverage probability conditioned on the relative distance in terms of the number of antennas per satellite, cluster size, satellite density, fading parameters, and the path-loss exponent. Our key finding is that coordinated beamforming is more beneficial than without satellite coordination when the number of satellite antennas is large and the relative in-cluster distance is small. We further derived the coverage probability averaged over the relative distance distribution. With this, we verified the average perspective of coordinated beamforming. We find that the optimal cluster size that maximizes the ergodic spectral efficiency increases with satellite density when the number of transmit antennas is sufficiently large.

An intriguing research direction involves examining the downlink coverage performance in scenarios where a ground station receiver employs interference cancellation techniques as in \cite{Lee2016}. Additionally, it would be valuable to explore the secrecy spectral efficiency of uplink satellite networks using the established $K$ nearest satellite distributions.


\begin{appendices}
\section{Proof of Lemma 1}
\begin{proof}
    The void probability where the number of visible satellites is zero in spherical cap $\mathcal{A}_r$ is computed as
    \begin{align}
        \mathbb{P}\left[\Phi(\mathcal{A}_r)=0\right] &= e^{-\lambda|\mathcal{A}_r|}=e^{-2\lambda\pi(R_{\sf S}-R_{\sf E}-h_r)R_{\sf S} }\nonumber\\ &=e^{-2\lambda \pi \left(R_{\sf S}-R_{\sf E}-\frac{(R_{\sf S}^2-R_{\sf E}^2)-r^2}{2R_{\sf E}}\right)R_{\sf S}}, \label{eq:void1}
    \end{align} 
    for $R_{\sf min} \le r \le R_{\sf max}$.
    
    On the other hand, the void probability where the number of points is zero in a circular ring $\mathcal{\tilde{A}}_r$ where the inner and outer radii are $R_{\sf min}$ and $r$ is given by
    \begin{align}
        &\mathbb{P}\left[\Phi(\mathcal{\tilde{A}}_r)\!=\!0\right] \!=\! e^{-\Tilde{\lambda}|\mathcal{\tilde{A}}_r|} \!=\! e^{-\Tilde{\lambda} \pi (r^2 - R_{\sf min}^2)} \overset{(a)}{=} e^{-\lambda \pi\frac{R_{\sf S}}{R_{\sf E}}(r^2 - R_{\sf min}^2)} \nonumber \\  &\! \overset{(b)}{=}\! e^{-\lambda \pi \frac{R_{\sf S}}{R_{\sf E}}(r^2-(R_{\sf S}-R_{\sf E})^2)} =e^{-2\lambda \pi \left(R_{\sf S}-R_{\sf E}-\frac{(R_{\sf S}^2-R_{\sf E}^2)-r^2}{2R_{\sf E}}\right)R_{\sf S}}, \label{eq:void2}
    \end{align}
    where (a) follows from $\Tilde{\lambda}=\frac{R_{\sf S}}{R_{\sf E}}\lambda$ and (b) follows from $R_{\sf min}=R_{\sf S}-R_{\sf E}$. By \eqref{eq:void1} and \eqref{eq:void2}, we verify that the two point processes have the same void probabilities. Since simple point processes are completely characterized by their void probabilities, they are the same point processes.
\end{proof}

\section{Proof of Lemma 2}

\begin{proof}
    First, we compute the probability of the number of visible satellites is between $1$ and $K-1$ as
\begin{align}
    &\mathbb{P}\left[1\le \Phi(\mathcal{A})\le K-1\right] = \sum_{n=1}^{K-1}\frac{\left(\Tilde{\lambda}|\tilde{\mathcal{A}}|\right)^n}{n!} e^{-\Tilde{\lambda}|\tilde{\mathcal{A}}|} \nonumber \\
    &=\sum_{n=1}^{K-1}\frac{(\Tilde{\lambda}\pi )^n\left(R_{\sf max}^2-R_{\sf min}^2\right)^n}{n!} e^{-\Tilde{\lambda}\pi\left(R_{\sf max}^2-R_{\sf min}^2\right)}.
\end{align}
To compute the $f_{\norm{\mathbf{d}_1}|1\le \Phi(\mathcal{A})\le K-1}(r_1)$, we compute the probability that the nearest neighbor of a point $\mathbf{x}_1$ is larger than $r_1$ conditioned that between $1$ and $K-1$ satellites exits in $\mathcal{A}$ as 
\begin{align}
    &\mathbb{P}\left[\norm{\mathbf{d}_1} > r_1 \big| 1\le\Phi(\mathcal{A})\le K-1\right] \nonumber\\&= \mathbb{P}\left[\Phi(\mathcal{A}_{r_1})=0 \big| 1\le\Phi(\mathcal{A})\le K-1\right] \nonumber \\ &=\frac{\mathbb{P}\left[\Phi(\mathcal{A}_{r_1})=0, 1\le\Phi(\mathcal{A})\le K-1\right]}{\mathbb{P}\left[1\le\Phi(\mathcal{A})\le K-1\right]} \nonumber \\ &\overset{(a)}{=}\frac{\mathbb{P}\left[\Phi(\mathcal{A}_{r_1})=0\right]\mathbb{P}\left[ 1\le\Phi(\mathcal{A}/\mathcal{A}_{r_1})\le K-1\right]}{\mathbb{P}\left[1\le\Phi(\mathcal{A})\le K-1\right]} \nonumber\\
    &=\frac{e^{-\Tilde{\lambda}|\Tilde{\mathcal{A}}_{r_1}|}\sum_{u=1}^{K-1}\frac{\left(\Tilde{\lambda}|\tilde{\mathcal{A}}/\tilde{\mathcal{A}}_{r_1}|\right)^u}{u!} e^{-\Tilde{\lambda}|\tilde{\mathcal{A}}/\tilde{\mathcal{A}}_{r_1}|}}{\sum_{n=1}^{K-1}\frac{\left(\Tilde{\lambda}|\tilde{\mathcal{A}}|\right)^n}{n!} e^{-\Tilde{\lambda}|\tilde{\mathcal{A}}|}},
\end{align}
for $R_{\sf min}\le r_1 \le R_{\sf max}$, where (a) is by the independence of the PPP for non-overlapping areas $\mathcal{A}$ and $\mathcal{A}/\mathcal{A}_r$. Since the area of $\mathcal{A}/\mathcal{A}_r$ is computed as $|\mathcal{A}/\mathcal{A}_r|= \pi(R_{\sf max}^2 -r^2)$, we obtain the conditional complementary CDF (CCDF) of $\norm{\mathbf{d}_1}$ as
\begin{align}
     F_{\norm{\mathbf{d}_1}|1\le\Phi(\mathcal{A})\le K-1}^c(r_1)=\frac{\sum_{u=1}^{K-1}(\Tilde{\lambda}\pi)^u\left(R_{\sf max}^2-r_1^2\right)^u/u!}{\sum_{n=1}^{K-1}( \Tilde{\lambda}\pi)^n\left(R_{\sf max}^2-R_{\sf min}^2\right)^n/n!}.
\end{align}
By taking derivative with respect to $r_1$, we obtain the conditional distribution of the nearest satellite distance as
\begin{align}
    &f_{\norm{\mathbf{d}_1}|1\le\Phi(\mathcal{A})\le K-1}(r_1)\nonumber\\&=\frac{\partial (1-F_{\norm{\mathbf{d}_1}|1\le\Phi(\mathcal{A})\le K-1}^c(r_1))}{\partial r_1} \nonumber\\&= \frac{\sum_{u=1}^{K-1}( \Tilde{\lambda}\pi)^u\left(R_{\sf max}^2-r_1^2\right)^{u-1}2r_1/(u-1)!}{\sum_{n=1}^{K-1}(\Tilde{\lambda}\pi)^n\left(R_{\sf max}^2-R_{\sf min}^2\right)^n/n!},
\end{align}
where $R_{\sf min} \le r_1\le R_{\sf max}$.
\end{proof}

\section{Proof of Lemma 3}
\begin{proof}
We compute the distribution of the $K$th nearest BS distance $||\mathbf{d}_K||$ conditioned that at least $K$ satellites are visible. 
Under the condition of $\delta$, the distance of the $K$th nearest BS is at least $\frac{R_{\sf min}}{\delta}$. Therefore, the conditional probability that the $K$th nearest neighbor of a point $\mathbf{x}_K$ is larger than $r_K$ is given by

\begin{align}
&\mathbb{P} \left[\norm{\mathbf{d}_K}>r_K \Big| \Phi(\mathcal{A})\ge K, \norm{\mathbf{d}_K} \ge \frac{R_{\sf min}}{\delta}\right]  \nonumber\\&= \sum_{i=0}^{K-1}\mathbb{P}\left[\Phi(\mathcal{A}_{r_K})=i \Big| \Phi(\mathcal{A})\ge K, \norm{\mathbf{d}_K} \ge \frac{R_{\sf min}}{\delta}\right], \label{eq:Kdistance}
\end{align}
where $\frac{R_{\sf min}}{\delta}\le r_K \le R_{\sf max}$.

To derive the conditional distribution of the $\norm{\mathbf{d}_K}$, we first compute the joint probability of the number of visible satellites is at least $K$ and the distance of the $K$th nearest BS is at least $\frac{R_{\sf min}}{\delta}$ as
\begin{align}
    &\mathbb{P}\left[\Phi(\mathcal{A})\ge K, \norm{\mathbf{d}_K} \ge \frac{R_{\sf min}}{\delta}\right]\nonumber \\&\overset{(a)}{=}\sum_{j=0}^{K-1}\mathbb{P}\left[\Phi(\mathcal{A}_{\frac{R_{\sf min}}{\delta}})=j\right] \mathbb{P}\left[\Phi(\mathcal{A}/\mathcal{A}_{\frac{R_{\sf min}}{\delta}}) \ge K-j\right] \nonumber \\
    &=\sum_{j=0}^{K-1}\frac{\left(\Tilde{\lambda}|\mathcal{A}_{\frac{R_{\sf min}}{\delta}}|\right)^je^{-\Tilde{\lambda} |\mathcal{A}_{\frac{R_{\sf min}}{\delta}}| }}{j!} 
    \nonumber\\ &~~~~~~~~\cdot\left(1-\sum_{w=0}^{K-j-1}\frac{ (\Tilde{\lambda}|\mathcal{A}/\mathcal{A}_{\frac{R_{\sf min}}{\delta}}|)^w e^{-\lambda|\mathcal{A}/\mathcal{A}_{\frac{R_{\sf min}}{\delta}}|}}{w!} \right) \nonumber \\
    &=\sum_{j=0}^{K-1}\frac{(\Tilde{\lambda}\pi)^j(\frac{R_{\sf min}^2}{\delta^2}-R_{\sf min}^2)^j}{j!}\Bigg(e^{-\Tilde{\lambda}\pi(\frac{R_{\sf min}^2}{\delta^2}-R_{\sf min}^2)}\nonumber\\ &-e^{-\Tilde{\lambda}\pi(R_{\sf max}^2-R_{\sf min}^2)}\sum_{w=0}^{K-j-1}\frac{(\Tilde{\lambda}\pi)^w(R_{\sf max}^2-\frac{R_{\sf min}^2}{\delta^2})^w}{w!}\Bigg),
\end{align}
where (a) is by the independence of the PPP for non-overlapping areas $\mathcal{A}_{\frac{R_{\sf min}}{\delta}}$ and $\mathcal{A}/\mathcal{A}_{\frac{R_{\sf min}}{\delta}}$.
Using this joint probability, the conditional probability inside the summation in \eqref{eq:Kdistance} is given by
\begin{align}
    &\mathbb{P}\left[\Phi(\mathcal{A}_{r_K})=i \Big| \Phi(\mathcal{A})\ge K, \norm{\mathbf{d}_K} \ge \frac{R_{\sf min}}{\delta}\right] \nonumber \\&\overset{(a)}{=} \frac{\mathbb{P}\left[\Phi(\mathcal{A}_{r_K})=i,\Phi(\mathcal{A})\ge K, \norm{\mathbf{d}_K} \ge \frac{R_{\sf min}}{\delta}\right]}{\mathbb{P}\left[\Phi(\mathcal{A})\ge K, \norm{\mathbf{d}_K} \ge \frac{R_{\sf min}}{\delta}\right]} \nonumber\\   &=\frac{\mathbb{P}\left[\Phi(\mathcal{A}_{r_K})=i\right]\mathbb{P}\left[\Phi(\mathcal{A}/\mathcal{A}_{r_K})\ge K-i\right]}{v(\Tilde{\lambda},R_{\sf min},R_{\sf max}, \delta,K)} \nonumber \\    &=\frac{\mathbb{P}\left[\Phi(\mathcal{A}_{r_K})=i\right]\left(1-\mathbb{P}\left[\Phi(\mathcal{A}/\mathcal{A}_{r_K})\le K-i-1\right]\right)}{v(\Tilde{\lambda},R_{\sf min},R_{\sf max}, \delta,K)} \nonumber \\ 
    &= \frac{\frac{\left(\Tilde{\lambda}|\mathcal{A}_{r_K}|\right)^i e^{-\Tilde{\lambda}|\mathcal{A}_{r_K}|}}{i!} \!\left(\!1\!-\!\sum_{t=0}^{K-i-1} \frac{(\tilde{\lambda}|\mathcal{A}/\mathcal{A}_{r_K}|)^t e^{-\tilde{\lambda}|\mathcal{A}/\mathcal{A}_{r_K}|}}{t!}\!\right)}{v(\Tilde{\lambda},R_{\sf min},R_{\sf max}, \delta,K)}\nonumber \\ &= \frac{\frac{(\Tilde{\lambda}\pi)^i(r_K^2-R_{\sf min}^2)^i}{i!}}{v(\Tilde{\lambda},R_{\sf min},R_{\sf max}, \delta,K)}\Bigg(e^{-\Tilde{\lambda}\pi (r_K^2-R_{\sf min}^2)}\nonumber\\ &~~~~~~~-e^{-\Tilde{\lambda}\pi(R_{\sf max}^2-R_{\sf min}^2)}\sum_{t=0}^{K-i-1}\frac{(\Tilde{\lambda}\pi)^t(R_{\sf max}^2-r_K^2)^t}{t!}\Bigg),
\end{align}
 where (a) is by the independence of the PPP for non-overlapping areas $\mathcal{A}$ and $\mathcal{A}/\mathcal{A}_{r_K}$.
Then, we obtain the conditional CCDF of $\norm{\mathbf{d}_K}$ as
\begin{align}
    &\mathbb{P} \left[\norm{\mathbf{d}_K}>r_K \Big| \Phi(\mathcal{A})\ge K, \norm{\mathbf{d}_K} \ge \frac{R_{\sf min}}{\delta}\right] \nonumber \\ & = \sum_{i=0}^{K-1}\mathbb{P}\left[\Phi(\mathcal{A}_{r_K})=i \Big| \Phi(\mathcal{A})\ge K, \norm{\mathbf{d}_K} \ge \frac{R_{\sf min}}{\delta}\right]  \nonumber\\&=\sum_{i=0}^{K-1} \frac{\frac{(\Tilde{\lambda}\pi)^i(r_K^2-R_{\sf min}^2)^i}{i!}}{v(\Tilde{\lambda},R_{\sf min},R_{\sf max}, \delta,K)}\Bigg(e^{-\Tilde{\lambda}\pi (r_K^2-R_{\sf min}^2)}\nonumber\\ &~~~~~~~-e^{-\Tilde{\lambda}\pi(R_{\sf max}^2-R_{\sf min}^2)}\sum_{t=0}^{K-i-1}\frac{(\Tilde{\lambda}\pi)^t(R_{\sf max}^2-r_K^2)^t}{t!}\Bigg)\nonumber\\  &= F_{\norm{\mathbf{d}_K} | \Phi(\mathcal{A}) \ge K,\delta}^c (r_K).
\end{align}
By taking derivative with respect to $r_K$, we obtain the conditional distribution of the $K$th nearest satellite distance as in \eqref{eq:K_nearest_dist}.
\end{proof}

\vspace{-0.3cm}
\section{Proof of Lemma 4}
\begin{proof}
   The conditional Laplace transform of the aggregated interference power plus noise power for $\norm{\mathbf{d}_K}=r_K$ is computed as
\begin{align}
    &\mathcal{L}_{I_{r_K}+\Bar{\sigma}^2|\Phi(\mathcal{A})\ge K}(s)\nonumber\\& = \mathbb{E}_{I_{r_K}}\bigg[e^{-s(I_{r_K}+\Bar{\sigma}^2)}\Big|\norm{\mathbf{d}_K}=r_K, \Phi(\mathcal{A})\ge K \bigg] \nonumber \\ &= \mathbb{E}_{I_{r_K}}\bigg[e^{-s \sum_{\mathbf{x}_n \in \Phi \cap \mathcal{A}_{r_K}^c}\Bar{G}_n H_n \norm{\mathbf{d}_n}^{-\alpha}} e^{-s \Bar{\sigma}^2}\nonumber\\&~~~~~~~~~~~~~~~~~~~~~~~~~~~~~~~~~~~~~\Big|\norm{\mathbf{d}_K}=r_K, \Phi(\mathcal{A})\ge K\bigg] \nonumber \\
    &\overset{(a)}{=} \mathbb{E}_{I_{r_K}}\Bigg[e^{-s\Bar{\sigma}^2}\prod_{\mathbf{x}_n \in \Phi \cap \mathcal{A}_{r_K}^c}e^{-s \Bar{G}_n H_n \norm{\mathbf{d}_n}^{-\alpha}}\nonumber\\&~~~~~~~~~~~~~~~~~~~~~~~~~~~~~~~~~~~~~\Bigg|\norm{\mathbf{d}_K}=r_K, \Phi(\mathcal{A})\ge K\Bigg] \nonumber \\
    &\overset{(b)}{=} e^{\left(-s\Bar{\sigma}^2  -\lambda \int_{v \in \mathcal{A}_{r_K}^c} \left(1-\mathbb{E}_{H_n}\left[e^{-s \Bar{G}_n H_n v^{-\alpha}}\right]\right)\mathrm{d}v\right)}\nonumber \\     
    &\overset{(c)}{=} e^{\left(-s\Bar{\sigma}^2  -2 \pi \tilde{\lambda}  \int_{r_K}^{R_{\sf max}}\left(1-\mathbb{E}_{H_n}\left[e^{-s \Bar{G}_n H_n v^{-\alpha}}\right]\right)v\mathrm{d}v \right)},  \label{eq:pf_Laplace}
\end{align}
where $\mathcal{A}_{r_K}^c=\mathcal{A}/\mathcal{A}_{r_K}$, (a) follows from the independence of $\mathbf{d}_n$ and $H_n$, (b) follows from the probability generating functional (PGFL) of the PPP \cite{Andrews2011}, and (c) come from $\frac{\partial |\mathcal{A}_{r_K}|}{\partial r_K}=2\frac{R_{\sf S}}{R_{\sf E}}\pi r_K$.

By assuming $m$ is integer, the PDF of $H_n$ is obtained as in \cite{An2016}
\begin{align}
    f_{H_n}(h) = \sum_{z=0}^{m-1} \zeta(z)h^z\exp\left(-(\beta-c)h\right).
\end{align}
Then, the inside expectation in \eqref{eq:pf_Laplace} is obtained by
\begin{align}
    &\mathbb{E}_{H_n}\left[e^{-s \Bar{G}_n H_n v^{-\alpha}}\right]  \nonumber\\&= \int_{0}^{\infty}e^{-s \Bar{G}_n h v^{-\alpha}} f_{H_n}(h) \mathrm{d}h \nonumber\\ &=\sum_{z=0}^{m-1} \zeta(z) \int_{0}^{\infty}e^{-s \Bar{G}_n h v^{-\alpha}}\zeta(z)h^z e^{\left(-(\beta-c)h\right)} \mathrm{d}h \nonumber\\ &= \sum_{z=0}^{m-1} \zeta(z) \frac{\Gamma(z+1)}{(\beta-c+s \Bar{G}_n v^{-\alpha})^{z+1}}.\label{eq:inside_exp}
\end{align}
Plugging \eqref{eq:inside_exp} into \eqref{eq:pf_Laplace}, we finally obtain the conditional Laplace transform of the aggregated interference power plus noise power in \eqref{eq:Laplace}.

\end{proof}

\section{Proof of Theorem 1}
\begin{proof}
    For $K\ge2$, the SINR coverage probability is composed by
\begin{align}
    &P_{{\sf SINR}|\delta}^{\sf Cov}(K, N_t, \alpha,\Tilde{\lambda}, R_{\sf min}, R_{\sf max}, \delta  ; \gamma) \nonumber\\& = \mathbb{P}\left[\Phi(\mathcal{A})=0\right] P_{{\sf SINR}|\Phi(\mathcal{A})= 0}^{\sf Cov}(K, N_t, \alpha,\Tilde{\lambda}, R_{\sf min}, R_{\sf max} ; \gamma)\nonumber \\&+ \mathbb{P}\left[1\le \Phi(\mathcal{A})\le K-1\right]\nonumber\\&~~~~~~~~~~~~~~\cdot P_{{\sf SINR}|1\le \Phi(\mathcal{A})\le K-1}^{\sf Cov}(K, N_t, \alpha,\Tilde{\lambda}, R_{\sf min}, R_{\sf max} ; \gamma) \nonumber\\ &+ \!\mathbb{P}\!\left[\Phi(\mathcal{A})\!\ge\! K\right] \!P_{{\sf SINR}|\Phi(\mathcal{A})\ge K,\delta}^{\sf Cov}(K, N_t, \alpha,\Tilde{\lambda}, R_{\sf min}, R_{\sf max},\delta ; \gamma \!).
\end{align}
Since $P_{{\sf SINR}|\Phi(\mathcal{A})= 0}^{\sf Cov}(K, N_t, \alpha,\Tilde{\lambda}, R_{\sf min}, R_{\sf max} ; \gamma)=0$, the conditional SINR coverage probability is finally obtained as
\begin{align}
    &P_{{\sf SINR}|\delta}^{\sf Cov}(K, N_t, \alpha,\Tilde{\lambda}, R_{\sf min}, R_{\sf max}, \delta  ; \gamma)\nonumber \\ &=\mathbb{P}\left[1\le \Phi(\mathcal{A})\le K-1\right] \nonumber\\ &~~~~~~~~~~~~~~\cdot P_{{\sf SINR}|1\le \Phi(\mathcal{A})\le K-1}^{\sf Cov}(K, N_t, \alpha,\Tilde{\lambda}, R_{\sf min}, R_{\sf max} ; \gamma) \nonumber\\ &+ \!\mathbb{P}\!\left[\Phi(\mathcal{A})\!\ge\! K\right] \!P_{{\sf SINR}|\Phi(\mathcal{A})\ge K,\delta}^{\sf Cov}(K, N_t, \alpha,\Tilde{\lambda}, R_{\sf min}, R_{\sf max},\delta ; \gamma \!). \label{eq:delta_Cov}
\end{align}
By assuming $m$ is integer, the CDF of $H_n$ is obtained as
\begin{align}
    &F_{H_n}(h) \nonumber\\&= \sum_{z=0}^{m-1} \frac{\zeta(z)}{(\beta-c)^{z+1}} \gamma(z+1,(\beta-c)h), \nonumber\\ &= \sum_{z=0}^{m-1} \frac{\zeta(z)}{(\beta-c)^{z+1}} \left(\Gamma(z)\!-\!\Gamma(z)e^{-(\beta-c)h}\sum_{v=0}^{z}\frac{(\beta-c)^v h^v}{v!}\right).
\end{align}

When $1\le \Phi(\mathcal{A})\le K-1$, we consider the SNR coverage since $K-1$ interfering signals are assumed to be eliminated with coordinated beamforming.
Then, we compute the conditional coverage probability conditioned on $1\le \Phi(\mathcal{A})\le K-1$ as
\begin{align}
    &P_{{\sf SINR}|1\le \Phi(\mathcal{A})\le K-1}^{\sf Cov}(K, N_t, \alpha,\Tilde{\lambda}, R_{\sf min}, R_{\sf max} ; \gamma) \nonumber \\ &=
    P_{{\sf SNR}|1\le \Phi(\mathcal{A})\le K-1}^{\sf Cov}(K, N_t, \alpha,\Tilde{\lambda}, R_{\sf min}, R_{\sf max} ; \gamma)
    \nonumber \\ &= \mathbb{E}_{r_1}\Bigg[\mathbb{P}\bigg[\frac{H_1 r_1^{-\alpha}}{\Bar{\sigma}^2}\ge \gamma \bigg|\norm{\mathbf{d}_1}=r_1, 1\le \Phi(\mathcal{A})\le K\bigg]\nonumber\\&~~~~~~~~~~~~~~~~~~~~~~~~~~~~~~~~~~~~~~~~~~~\Bigg|1\le \Phi(\mathcal{A})\le K-1\Bigg] \nonumber \\  &= \mathbb{E}_{r_1}\Bigg[1-\sum_{z=0}^{m-1}\frac{\zeta(z)}{(\beta-c)^{z+1}}\Bigg(\Gamma(z)-\Gamma(z)
e^{-(\beta-c)\gamma r_1^{\alpha}\Bar{\sigma}^2}\nonumber\\&~~~~~~~~~~~~~~ \cdot\sum_{v=0}^{z}\frac{((\beta-c)\gamma r_1^\alpha \Bar{\sigma}^2)^v}{v!}\Bigg) \Bigg|1\le \Phi(\mathcal{A})\le K-1\Bigg] \nonumber \\ 
    &= \!\int_{R_{\sf min}}^{R_{\sf max}} \Bigg[1-\sum_{z=0}^{m-1}\frac{\zeta(z)\Gamma(z)}{(\beta-c)^{z+1}}\Bigg(1-
e^{-(\beta-c)\gamma r_1^{\alpha}\Bar{\sigma}^2}\nonumber\\&~\cdot\sum_{v=0}^{z}\frac{((\beta-c)\gamma r_1^\alpha \Bar{\sigma}^2)^v}{v!}\Bigg) \Bigg] f_{\norm{\mathbf{d}_1}|1 \! \le \! \Phi(\mathcal{A})\le K-1}(r_1)\mathrm{d}r_1. \label{eq:cond_covK-1}
\end{align}

Further, the coverage probability conditioned on $\Phi(\mathcal{A})\ge K$ is given by
\begin{align}
    &P_{{\sf SINR}|\Phi(\mathcal{A})\ge K,\delta}^{\sf Cov}(K, N_t,, \alpha,\Tilde{\lambda}, R_{\sf min}, R_{\sf max},\delta; \gamma)\nonumber \\ &= \mathbb{E}_{r_K}\Bigg[\mathbb{P}\bigg[\frac{H_1 \delta ^{-\alpha}r_K^{-\alpha}}{I_{r_K}+\Bar{\sigma}^2}\ge \gamma \bigg|\norm{\mathbf{d}_K}=r_K, \Phi(\mathcal{A})\ge K,\delta\bigg]\nonumber\\&~~~~~~~~~~~~~~~~~~~~~~~~~~~~~~~~~~~~~~~~~~~~~~~~~~~~~~\Bigg|\Phi(\mathcal{A})\ge K,\delta\Bigg] \nonumber \\  &= \mathbb{E}_{r_K}\bigg[\mathbb{P}\Big[H_1 \ge \gamma (\delta r_K)^{\alpha}(I_{r_K}+\Bar{\sigma}^2) \nonumber\\&~~~~~~~~~~~~~~~~~~~~\Big|\norm{\mathbf{d}_K}=r_K, \Phi(\mathcal{A})\ge K,\delta\Big]\bigg|\Phi(\mathcal{A})\ge K,\delta\bigg]\nonumber\\ &=\mathbb{E}_{r_K}\!\Bigg[\mathbb{E}_{I_{r_K}}\!\Bigg[1\!\!-\!\!\sum_{z=0}^{m-1}\frac{\zeta(z)\Gamma(z)}{(\beta\!-\!c)^{z+1}}\Bigg(1 \!-\!\sum_{v=0}^{z} \frac{((\beta\!-\!c)\gamma(\delta r_K)^\alpha)^v}{v!} \nonumber\\&~~~~~~ \cdot(I_{r_k}+\Bar{\sigma}^2)^v e^{-(\beta-c)\gamma(\delta r_K)^\alpha (I_{r_k}+\Bar{\sigma}^2)}\Bigg)\Bigg]\!\Bigg|\Phi(\mathcal{A})\!\ge\! K,\delta\!\Bigg] \nonumber \\ 
    &\overset{(a)}{=} \mathbb{E}_{r_K}\Bigg[1\!\!-\!\!\sum_{z=0}^{m-1}\frac{\zeta(z)\Gamma(z)}{(\beta\!-\!c)^{z+1}}\Bigg(\!\!1 \!-\!\sum_{v=0}^{z} \frac{((\beta\!-\!c)\gamma(\delta r_K)^\alpha)^v}{v!}(-1)^v \nonumber\\&~~~\cdot\frac{\mathrm{d}^v \mathcal{L}_{I_{r_K}+\Bar{\sigma}^2|\Phi(\mathcal{A})\ge K}(s)}{\mathrm{d}s^v} \bigg|_{s=(\beta-c)(\delta r_K)^{\alpha}\gamma} \Bigg)\Bigg| \Phi(\mathcal{A})\ge K,\delta\Bigg] \nonumber \\ 
    &\overset{(b)}{=} \int_{\frac{R_{\sf min}}{\delta}}^{R_{\sf max}} \Bigg[1\!\!-\!\!\sum_{z=0}^{m-1}\frac{\zeta(z)\Gamma(z)}{(\beta\!-\!c)^{z+1}}\Bigg(1 \!-\!\sum_{v=0}^{z} \frac{((\beta\!-\!c)\gamma(\delta r_K)^\alpha)^v}{v!} \nonumber\\&~~~~~~~~~~~~\cdot(-1)^v\frac{\mathrm{d}^v \mathcal{L}_{I_{r_K}+\Bar{\sigma}^2|\Phi(\mathcal{A})\ge K}(s)}{\mathrm{d}s^v} \bigg|_{s=(\beta-c)(\delta r_K)^{\alpha}\gamma} \Bigg)\Bigg] \nonumber\\ &~~~~~~~~~~~~~~~~~~~~~~~~~~~~~~\cdot f_{\norm{\mathbf{d}_K} | \Phi(\mathcal{A}) \ge K, \delta} (r_K)  \mathrm{d}r_K, \label{eq:cond_covK}
\end{align}
where (a) follows from applying the derivative property of the Laplace transform, i.e., $\mathbb{E}\left[X^v e^{-sX}\right]=(-1)^v \frac{\mathrm{d}\mathcal{L}_X(s)}{\mathrm{d}s^v}$, and (b) comes from the expectation over the $K$th nearest distribution obtained in Lemma \ref{lem:Knearest}.

By combining \eqref{eq:cond_covK-1} and \eqref{eq:cond_covK}, we finally obtain the conditional SINR coverage probability in \eqref{eq:theorem1}, which completes the proof.
\end{proof}

\section{Proof of Theorem 2}
\begin{proof}
From Alzer's inequality \cite{alzer1997}, if $\sqrt{H_1}$ is Shadowed-Rician distributed, the CCDF of $H_1$ is upper and lower bounded by

    \begin{align}
    &1-\sum_{z=0}^{m-1}\frac{\zeta(z)}{(\beta-c)^{z+1}}z!\left(1-e^{-\kappa^{\sf L}(\beta-c)h}\right)^{z+1} \le \mathbb{P}[H_1 > h] \nonumber\\ &\le 1-\sum_{z=0}^{m-1}\frac{\zeta(z)}{(\beta-c)^{z+1}}z!\left(1-e^{-\kappa^{\sf U}(\beta-c)h}\right)^{z+1}, 
\end{align}
where $\kappa^{\sf L}=1$ and $\kappa^{\sf U}=\left((z+1)!\right)^{-\frac{1}{z+1}}$. Then, by adjusting parameter $\kappa$ between $\kappa^{\sf L}$ and $\kappa^{\sf U}$, the CCDF of $H_1$ is approximated as
\begin{align}
    \mathbb{P}[H_1 > h] \simeq 1-\sum_{z=0}^{m-1}\frac{\zeta(z)}{(\beta-c)^{z+1}}z!\left(1-e^{-\kappa(\beta-c)h}\right)^{z+1},\nonumber
\end{align}
where $ \left((z+1)!\right)^{-\frac{1}{z+1}}\le \kappa \le 1$ and the equality holds when $m=1$. Then, By applying the binomial expansion as
\begin{align}
    \left(1-e^{-\kappa(\beta-c)h}\right)^{z+1} = \sum_{\ell = 0}^{z+1} \binom{z+1}{\ell}(-1)^{\ell}e^{-\ell \kappa (\beta-c)x},
\end{align}
we obtain an approximation of the conditional CCDF of SINR by
\begin{align}
   & \mathbb{P}\left[H_1 \ge \gamma (\delta r_K)^{\alpha}(I_{r_K}+\Bar{\sigma}^2) \Big|\norm{\mathbf{d}_K}=r_K, \Phi(\mathcal{A})\ge K,\delta\right] \nonumber \\
   & \simeq 1-\sum_{z=0}^{m-1}\frac{\zeta(z)}{(\beta-c)^{z+1}}z!\sum_{\ell = 0}^{z+1} \binom{z+1}{\ell}(-1)^{\ell} \nonumber\\& \cdot \mathbb{E}_{I_{r_K}}\Big[e^{-\ell \kappa (\beta\!-\!c)\gamma (\delta r_K)^{\alpha}(I_{r_K}+\Bar{\sigma}^2) }\Big| \!\norm{\mathbf{d}_K}=r_K, \Phi(\mathcal{A})\!\ge\! K,\delta \Big].
\end{align}
Finally, we obtain the approximation of the conditional coverage probability by
\begin{align}
    & P_{{\sf SINR}|\Phi(\mathcal{A})\ge K,\delta}^{\sf Cov}(K, N_t, \alpha,\Tilde{\lambda}, R_{\sf min}, R_{\sf max},\delta; \gamma) \nonumber \\
    &\simeq 1\!-\sum_{z=0}^{m-1}\frac{\zeta(z)}{(\beta-c)^{z+1}}z!\sum_{\ell = 0}^{z+1}\! \binom{z\!+\!1}{\ell}(-1)^{\ell} \nonumber\\ &~\cdot\mathbb{E}_{r_K} \bigg[\mathbb{E}_{I_{r_K}}\Big[e^{-\ell \kappa (\beta\!-\!c)\gamma (\delta r_K)^{\alpha}(I_{r_K}+\Bar{\sigma}^2) }\nonumber\\&~~~~~~~~~~~~~~~~~~\Big| \norm{\mathbf{d}_K}=r_K, \Phi(\mathcal{A})\ge K,\delta \Big]\bigg|\Phi(\mathcal{A})\ge K, \delta \bigg] \nonumber \\
    &=1-\sum_{z=0}^{m-1}\frac{\zeta(z)}{(\beta-c)^{z+1}}z!\sum_{\ell = 0}^{z+1} \binom{z+1}{\ell}(-1)^{\ell} \nonumber\\&\cdot\mathbb{E}_{r_K} \Big[\mathcal{L}_{I_{r_K}+\Bar{\sigma}^2|\Phi(\mathcal{A})\ge K}\left(\ell  \kappa (\beta\!-\!c)\gamma (\delta r_K)^{\alpha}\right)\Big|\Phi(\mathcal{A})\ge K, \delta \Big].
\end{align} 
\end{proof}

\section{Proof of Lemma 5}
\begin{proof}
    To derive the conditional distribution of $\delta$, we begin by computing the joint PDF of $\norm{\mathbf{d}_1}$, $\norm{\mathbf{d}_K}$, and $\Phi(\mathcal{A})\ge K$. The joint PDF is obtained by considering the four non-overlapping areas $A_1=\mathcal{A}_{r_1}/\mathcal{A}_{R_{\sf min}}$, $A_2=\mathcal{A}_{r_1+\mathrm{d}r_1}/\mathcal{A}_{r_1}$, $A_3=\mathcal{A}_{r_K}/\mathcal{A}_{r_1+\mathrm{d}r_1}$, and $A_4=\mathcal{A}_{r_K+\mathrm{d}r_K}/\mathcal{A}_{r_K}$ in the replaced $\mathbb{R}^2$ circular ring. Specifically, the joint probability that $\norm{\mathbf{d}_1}$ belongs to area $A_2$, $\norm{\mathbf{d}_K}$ belongs to area $A_4$, and $\Phi(\mathcal{A})\ge K$ is given by the product of the four independent probability events as
    \begin{align}
        &\mathbb{P}\left[\norm{\mathbf{d}_1}\in A_2, \norm{\mathbf{d}_K}\in A_4, \Phi(\mathcal{A})\ge K \right] \nonumber\\&=  \begin{cases}
      P_1 P_2 P_3 P_4  &  \text{if}~ R_{\sf mim}\le r_1 \le r_K \le R_{\sf max} \\
      0 &  \text{otherwise}
    \end{cases},
    \end{align}
    where
    \begin{align}
    P_1  &= \mathbb{P}\left[\text{No points in}~ A_1\right] = e^{-\Tilde{\lambda}\pi(r_1^2- R_{\sf min}^2)} \nonumber \\ P_2 &= \mathbb{P}\left[\text{One point in}~ A_2\right] =\Tilde{\lambda}\pi 2 r_1 \mathrm{d}r_1 e^{-\Tilde{\lambda} \pi 2 r_1 \mathrm{d}r_1} \nonumber \\ 
    P_3 & = \mathbb{P}\left[\text{$K-2$ points in}~ A_3\right] \nonumber\\&= \frac{(\tilde{\lambda}\pi)^{K-2}}{(K-2)!} \left[r_K^2\!-\!(r_1+\mathrm{d}r_1)^2\right]^{K-2} e^{- \Tilde{\lambda}\pi \left[r_K^2-(r_1+\mathrm{d}r_1)^2\right]} \nonumber \\
    P_4 & = \mathbb{P}\left[\text{One point in}~ A_4\right] =\Tilde{\lambda}\pi 2 r_K \mathrm{d}r_K e^{-\Tilde{\lambda} \pi 2 r_K \mathrm{d}r_K}.
    \end{align}
Then, the conditional joint probability is given by
\begin{align}
    &\mathbb{P}\left[\norm{\mathbf{d}_1}\in A_2, \norm{\mathbf{d}_K}\in A_4 \big| \Phi(\mathcal{A})\ge K \right]  \nonumber\\&=\begin{cases}
      \frac{P_1 P_2 P_3 P_4}{\mathbb{P}\left[\Phi(\mathcal{A})\ge K\right]}  &  \text{if}~ R_{\sf mim}\le r_1 \le r_K \le R_{\sf max} \\
      0 &  \text{otherwise}
    \end{cases}. \label{eq:condjointPDF}
\end{align}
The conditional joint PDF of $\norm{\mathbf{d}_1}$ and $\norm{\mathbf{d}_K}$ is obtained as a result of evaluating the limits of the conditional joint probability in \eqref{eq:condjointPDF}, which is given by 
\begin{align}
    &f_{\norm{\mathbf{d}_1},\norm{\mathbf{d}_K}|\Phi(\mathcal{A})\ge K} (r_1,r_K) \nonumber\\&=\lim_{\mathrm{d}r_1,\mathrm{d}r_K \rightarrow 0} \frac{\mathbb{P}\left[\norm{\mathbf{d}_1}\in A_2, \norm{\mathbf{d}_K}\in A_4 \big| \Phi(\mathcal{A})\ge K \right]}{\mathrm{d}r_1\mathrm{d}r_K} \nonumber\\&= 
        \frac{4\left(\Tilde{\lambda}\pi\right)^K r_1 r_K(r_K^2-r_1^2)^{K-2}e^{-\Tilde{\lambda}\pi(r_K^2-R_{\sf min}^2)}}{(K-2)!\mathbb{P}[\Phi(\mathcal{A})\ge K]},
\end{align}
for $R_{\sf mim}\le r_1 \le r_K \le R_{\sf max}$.
Using this conditional joint PDF, we obtain the CDF of $\delta$ as
\begin{align}
    &F_{\delta|\Phi(\mathcal{A})\ge K}(x)=\mathbb{P}[\delta\le x|\Phi(\mathcal{A})\ge K] = \mathbb{P}\left[\frac{\norm{\mathbf{d}_1}}{\norm{\mathbf{d}_K}}\le x\right] \nonumber\\&= \mathbb{P}\left[\norm{\mathbf{d}_1}\le x\norm{\mathbf{d}_K}\right] \nonumber\\ &=\int_{\frac{R_{\sf min}}{x}}^{R_{\sf max}} \int_{R_{\sf min}}^{x r_K}f_{\norm{\mathbf{d}_1},\norm{\mathbf{d}_K}|\Phi(\mathcal{A})\ge K} (r_1,r_K) \mathrm{d}r_1 \mathrm{d}r_K  \nonumber\\ &= \!\!\!\int_{\frac{R_{\sf min}}{x}}^{R_{\sf max}}\!\! \!\int_{R_{\sf min}}^{x r_K}\! \!\frac{4(\!\Tilde{\lambda}\pi\!
    )^K r_1 r_K(r_K^2\!-\!r_1^2)^{K\!-\!2}e^{-\Tilde{\lambda}\pi(r_K^2\!-\!R_{\sf min}^2)}}{(K-2)!\mathbb{P}[\Phi(\mathcal{A})\ge K]} \mathrm{d}r_1 \mathrm{d}r_K \nonumber \\ 
    &=\frac{1}{(K\!-\!1)!\left(1\!-\!\sum_{n=0}^{K-1}\frac{(\Tilde{\lambda}\pi )^n\left(R_{\sf max}^2-R_{\sf min}^2\right)^n}{n!} e^{-\Tilde{\lambda}\pi\left(R_{\sf max}^2-R_{\sf min}^2\right)}\right)}\nonumber\\&\cdot\Bigg(\int_{\Tilde{\lambda}\pi\left(\frac{1}{x^2}-1\right)R_{\sf min}^2}^{\Tilde{\lambda}\pi(R_{\sf max}^2-R_{\sf min}^2)}e^{-u}u^{K-1}\mathrm{d}u \nonumber\\&~~~~~~~~~~-e^{\Tilde{\lambda}\pi R_{\sf min}^2}(1-x^2)^{K-1} \int_{\Tilde{\lambda}\pi \frac{R_{\sf min}^2}{x^2}}^{\Tilde{\lambda}\pi R_{\sf max}^2}e^{-t}t^{K-1}\mathrm{d}t \Bigg),
\end{align}
where $\frac{R_{\sf min}}{R_{\sf max}}\le x \le 1$. By taking derivative with respect to $x$, we obtain the conditional distribution of $\delta$ as in \eqref{eq:delta_dist_lemma}.
\end{proof}

\section{Proof of Theorem 3}
\begin{proof}
    The SINR coverage probability is obtained by marginalizing the conditional coverage probability in Theorem 1 with respect to $\delta$ as
    \begin{align}
        &P_{{\sf SINR}}^{\sf Cov}(K, N_t, \alpha,\Tilde{\lambda}, R_{\sf min}, R_{\sf max}  ; \gamma) \nonumber\\&=\mathbb{E}_{\delta}\left[P_{{\sf SINR}|\delta}^{\sf Cov}(K, N_t, \alpha,\Tilde{\lambda}, R_{\sf min}, R_{\sf max}, \delta  ; \gamma) \right] \nonumber\\ &=\mathbb{P}\left[1\le \Phi(\mathcal{A})\le K-1\right] \nonumber\\ &\cdot \! P_{{\sf SINR}|1\le \Phi(\mathcal{A})\le K\!-\!1}^{\sf Cov}(K,\! N_t,\! \alpha,\!\Tilde{\lambda}, R_{\sf min}, R_{\sf max} ;\! \gamma) \!+ \!\mathbb{P}\!\left[\Phi(\mathcal{A})\!\ge\! K\right] \nonumber\\ &\cdot \int_{\frac{R_{\sf min}}{R_{\sf max}}}^{1} P_{{\sf SINR}|\Phi(\mathcal{A})\ge K,x}^{\sf Cov}(K, N_t, \alpha,\Tilde{\lambda}, R_{\sf min}, R_{\sf max},x ; \gamma \!) \nonumber\\ & ~~~~~~~~~~~~~~~~~~~~~~~\cdot f_{\delta|\Phi(\mathcal{A})\ge K}(x) \mathrm{d}x.
    \end{align}
\end{proof}

\end{appendices}

\bibliographystyle{IEEEtran}
\bibliography{IEEEabrv,Reference}

\begin{thebibliography}{10}
\providecommand{\url}[1]{#1}
\csname url@samestyle\endcsname
\providecommand{\newblock}{\relax}
\providecommand{\bibinfo}[2]{#2}
\providecommand{\BIBentrySTDinterwordspacing}{\spaceskip=0pt\relax}
\providecommand{\BIBentryALTinterwordstretchfactor}{4}
\providecommand{\BIBentryALTinterwordspacing}{\spaceskip=\fontdimen2\font plus
\BIBentryALTinterwordstretchfactor\fontdimen3\font minus \fontdimen4\font\relax}
\providecommand{\BIBforeignlanguage}[2]{{%
\expandafter\ifx\csname l@#1\endcsname\relax
\typeout{** WARNING: IEEEtran.bst: No hyphenation pattern has been}%
\typeout{** loaded for the language `#1'. Using the pattern for}%
\typeout{** the default language instead.}%
\else
\language=\csname l@#1\endcsname
\fi
#2}}
\providecommand{\BIBdecl}{\relax}
\BIBdecl

\bibitem{Zhu2022}
X.~Zhu and C.~Jiang, ``Integrated satellite-terrestrial networks toward {6G}: Architectures, applications, and challenges,'' \emph{{IEEE} Internet Things J.}, vol.~9, no.~1, pp. 437--461, Jan. 2022.

\bibitem{Liu2021}
S.~Liu, Z.~Gao, Y.~Wu, D.~W. Kwan~Ng, X.~Gao, K.-K. Wong, S.~Chatzinotas, and B.~Ottersten, ``{LEO} satellite constellations for {5G} and beyond: How will they reshape vertical domains?'' \emph{{IEEE} Commun. Mag.}, vol.~59, no.~7, pp. 30--36, Jul. 2021.

\bibitem{Del2019}
I.~Del~Portillo, B.~G. Cameron, and E.~F. Crawley, ``A technical comparison of three low earth orbit satellite constellation systems to provide global broadband,'' \emph{Acta astronautica}, vol. 159, pp. 123--135, 2019.

\bibitem{Park2023}
J.~Park, J.~Choi, and N.~Lee, ``A tractable approach to coverage analysis in downlink satellite networks,'' \emph{IEEE Transactions on Wireless Communications}, vol.~22, no.~2, pp. 793--807, Feb. 2023.

\bibitem{Andrews2011}
J.~G. Andrews, F.~Baccelli, and R.~K. Ganti, ``A tractable approach to coverage and rate in cellular networks,'' \emph{{IEEE} Trans. Commun.}, vol.~59, no.~11, pp. 3122--3134, Nov. 2011.

\bibitem{Su2019}
Y.~Su, Y.~Liu, Y.~Zhou, J.~Yuan, H.~Cao, and J.~Shi, ``Broadband {LEO} satellite communications: Architectures and key technologies,'' \emph{{IEEE} Wireless Commun.}, vol.~26, no.~2, pp. 55--61, Apr. 2019.

\bibitem{Clerckx2013}
B.~Clerckx, H.~Lee, Y.-J. Hong, and G.~Kim, ``A practical cooperative multicell {MIMO-OFDMA} network based on rank coordination,'' \emph{{IEEE} Trans. Wireless Commun.}, vol.~12, no.~4, pp. 1481--1491, 2013.

\bibitem{Lee2012}
D.~Lee, H.~Seo, B.~Clerckx, E.~Hardouin, D.~Mazzarese, S.~Nagata, and K.~Sayana, ``Coordinated multipoint transmission and reception in {LTE}-advanced: deployment scenarios and operational challenges,'' \emph{{IEEE} Commun. Mag.}, vol.~50, no.~2, pp. 148--155, 2012.

\bibitem{Lee2015-2}
N.~Lee, D.~Morales-Jimenez, A.~Lozano, and R.~W. Heath, ``Spectral efficiency of dynamic coordinated beamforming: A stochastic geometry approach,'' \emph{{IEEE} Trans. Wireless Commun.}, vol.~14, no.~1, pp. 230--241, Jan. 2015.

\bibitem{Baccelli2009}
F.~Baccelli and B.~B{\l}aszczyszyn, ``Stochastic geometry and wireless networks: Volume i theory,'' \emph{Found. Trends in Networking}, vol.~3, no. 3-4, pp. 249--449, Mar. 2009.

\bibitem{Haenggi2008}
M.~Haenggi, ``A geometric interpretation of fading in wireless networks: Theory and applications,'' \emph{{IEEE} Trans. Inf. Theory}, vol.~54, no.~12, pp. 5500--5510, Dec. 2008.

\bibitem{Guo2013}
A.~Guo and M.~Haenggi, ``Spatial stochastic models and metrics for the structure of base stations in cellular networks,'' \emph{{IEEE} Trans. Wireless Commun.}, vol.~12, no.~11, pp. 5800--5812, Nov. 2013.

\bibitem{Dhillon2012}
H.~S. Dhillon, R.~K. Ganti, F.~Baccelli, and J.~G. Andrews, ``Modeling and analysis of {K}-tier downlink heterogeneous cellular networks,'' \emph{{IEEE} J. Sel. Areas Commun.}, vol.~30, no.~3, pp. 550--560, Apr. 2012.

\bibitem{Heath2013}
R.~W. Heath, M.~Kountouris, and T.~Bai, ``Modeling heterogeneous network interference using poisson point processes,'' \emph{{IEEE} Trans. Signal Process.}, vol.~61, no.~16, pp. 4114--4126, Aug. 2013.

\bibitem{ElSawy2014}
H.~ElSawy and E.~Hossain, ``On stochastic geometry modeling of cellular uplink transmission with truncated channel inversion power control,'' \emph{{IEEE} Trans. Wireless Commun.}, vol.~13, no.~8, pp. 4454--4469, Aug. 2014.

\bibitem{Di2016}
M.~Di~Renzo and P.~Guan, ``Stochastic geometry modeling and system-level analysis of uplink heterogeneous cellular networks with multi-antenna base stations,'' \emph{{IEEE} Trans. Commun.}, vol.~64, no.~6, pp. 2453--2476, Jun. 2016.

\bibitem{Novlan2013}
T.~D. Novlan, H.~S. Dhillon, and J.~G. Andrews, ``Analytical modeling of uplink cellular networks,'' \emph{{IEEE} Trans. Wireless Commun.}, vol.~12, no.~6, pp. 2669--2679, Jun. 2013.

\bibitem{Ganti2012}
R.~K. Ganti and M.~Haenggi, ``Spatial analysis of opportunistic downlink relaying in a two-hop cellular system,'' \emph{{IEEE} Trans. Commun.}, vol.~60, no.~5, pp. 1443--1450, May 2012.

\bibitem{Elkotby2015}
H.~Elkotby and M.~Vu, ``Uplink user-assisted relaying in cellular networks,'' \emph{{IEEE} Trans. Wireless Commun.}, vol.~14, no.~10, pp. 5468--5483, Oct. 2015.

\bibitem{Lee2015}
N.~Lee, X.~Lin, J.~G. Andrews, and R.~W. Heath, ``Power control for {D2D} underlaid cellular networks: Modeling, algorithms, and analysis,'' \emph{{IEEE} J. Sel. Areas Commun.}, vol.~33, no.~1, pp. 1--13, Jan. 2015.

\bibitem{Park2016}
J.~Park, N.~Lee, J.~G. Andrews, and R.~W. Heath, ``On the optimal feedback rate in interference-limited multi-antenna cellular systems,'' \emph{{IEEE} Trans. Wireless Commun.}, vol.~15, no.~8, pp. 5748--5762, Aug. 2016.

\bibitem{Tanbourgi2015}
R.~Tanbourgi, H.~S. Dhillon, and F.~K. Jondral, ``Analysis of joint transmit–receive diversity in downlink {MIMO} heterogeneous cellular networks,'' \emph{{IEEE} Trans. Wireless Commun.}, vol.~14, no.~12, pp. 6695--6709, Dec. 2015.

\bibitem{Bai2015}
T.~Bai and R.~W. Heath, ``Coverage and rate analysis for millimeter-wave cellular networks,'' \emph{{IEEE} Trans. Wireless Commun.}, vol.~14, no.~2, pp. 1100--1114, Feb. 2015.

\bibitem{Singh2015}
S.~Singh, M.~N. Kulkarni, A.~Ghosh, and J.~G. Andrews, ``Tractable model for rate in self-backhauled millimeter wave cellular networks,'' \emph{{IEEE} J. Sel. Areas Commun.}, vol.~33, no.~10, pp. 2196--2211, Oct. 2015.

\bibitem{Di2015}
M.~Di~Renzo, ``Stochastic geometry modeling and analysis of multi-tier millimeter wave cellular networks,'' \emph{{IEEE} Trans. Wireless Commun.}, vol.~14, no.~9, pp. 5038--5057, Sept. 2015.

\bibitem{Chetlur2017}
V.~V. Chetlur and H.~S. Dhillon, ``Downlink coverage analysis for a finite {3-D} wireless network of unmanned aerial vehicles,'' \emph{IEEE Transactions on Communications}, vol.~65, no.~10, pp. 4543--4558, Oct. 2017.

\bibitem{Alkama2022}
D.~Alkama, M.~A. Ouamri, M.~S. Alzaidi, R.~N. Shaw, M.~Azni, and S.~S.~M. Ghoneim, ``Downlink performance analysis in {MIMO} {UAV}-cellular communication with {LOS/NLOS} propagation under {3D} beamforming,'' \emph{IEEE Access}, vol.~10, pp. 6650--6659, 2022.

\bibitem{Okati2020}
N.~Okati, T.~Riihonen, D.~Korpi, I.~Angervuori, and R.~Wichman, ``Downlink coverage and rate analysis of low earth orbit satellite constellations using stochastic geometry,'' \emph{{IEEE} Trans. Commun.}, vol.~68, no.~8, pp. 5120--5134, Aug. 2020.

\bibitem{Talgat2020}
A.~Talgat, M.~A. Kishk, and M.-S. Alouini, ``Nearest neighbor and contact distance distribution for binomial point process on spherical surfaces,'' \emph{{IEEE} Commun. Lett.}, vol.~24, no.~12, pp. 2659--2663, Dec. 2020.

\bibitem{Talgat2021}
------, ``Stochastic geometry-based analysis of leo satellite communication systems,'' \emph{{IEEE} Commun. Lett.}, vol.~25, no.~8, pp. 2458--2462, Aug. 2021.

\bibitem{Okati2022}
N.~Okati and T.~Riihonen, ``Nonhomogeneous stochastic geometry analysis of massive leo communication constellations,'' \emph{{IEEE} Trans. Commun.}, vol.~70, no.~3, pp. 1848--1860, Mar. 2022.

\bibitem{Al2021}
A.~Al-Hourani, ``An analytic approach for modeling the coverage performance of dense satellite networks,'' \emph{{IEEE} Wireless Commun. Lett.}, vol.~10, no.~4, pp. 897--901, Apr. 2021.

\bibitem{Chae2023}
S.~H. Chae, H.~Lim, H.~Lee, and B.~C. Jung, ``Performance analysis of dense low earth orbit satellite communication networks with stochastic geometry,'' \emph{J.Commn.Net}, vol.~25, no.~2, pp. 208--221, Apr. 2023.

\bibitem{park2023_2}
J.~Park, J.~Choi, N.~Lee, and F.~Baccelli, ``Unified modeling and rate coverage analysis for satellite-terrestrial integrated networks: Coverage extension or data offloading?'' \emph{arXiv preprint arXiv:2307.03343}, 2023.

\bibitem{lee2022}
J.~Lee, S.~Noh, S.~Jeong, and N.~Lee, ``Coverage analysis of leo satellite downlink networks: Orbit geometry dependent approach,'' \emph{arXiv preprint arXiv:2206.09382}, 2022.

\bibitem{Haenggi2005}
M.~Haenggi, ``On distances in uniformly random networks,'' \emph{{IEEE} Trans. Inf. Theory}, vol.~51, no.~10, pp. 3584--3586, Oct. 2005.

\bibitem{cundy1989}
H.~Cundy and A.~Rollett, ``Sphere and cylinder----archimedes’ theorem,'' \emph{Mathematical Models}, pp. 172--173, 1989.

\bibitem{Li2003}
J.~Li, P.~Stoica, and Z.~Wang, ``On robust capon beamforming and diagonal loading,'' \emph{{IEEE} Trans. Signal Process.}, vol.~51, no.~7, pp. 1702--1715, Jul. 2003.

\bibitem{Bai2014}
T.~Bai, A.~Alkhateeb, and R.~W. Heath, ``Coverage and capacity of millimeter-wave cellular networks,'' \emph{{IEEE} Commun. Mag.}, vol.~52, no.~9, pp. 70--77, Sept. 2014.

\bibitem{Renzo2015}
M.~Di~Renzo, ``Stochastic geometry modeling and analysis of multi-tier millimeter wave cellular networks,'' \emph{{IEEE} Trans. Wireless Commun.}, vol.~14, no.~9, pp. 5038--5057, Sept. 2015.

\bibitem{Abdi2003}
A.~Abdi, W.~Lau, M.-S. Alouini, and M.~Kaveh, ``A new simple model for land mobile satellite channels: first- and second-order statistics,'' \emph{{IEEE} Trans. Wireless Commun.}, vol.~2, no.~3, pp. 519--528, May 2003.

\bibitem{magnus1967}
W.~Magnus, F.~Oberhettinger, R.~P. Soni, and E.~P. Wigner, ``Formulas and theorems for the special functions of mathematical physics,'' \emph{Physics Today}, vol.~20, no.~12, pp. 81--83, 1967.

\bibitem{gradshteyn2014}
I.~S. Gradshteyn and I.~M. Ryzhik, \emph{Table of integrals, series, and products}.\hskip 1em plus 0.5em minus 0.4em\relax Academic press, 2014.

\bibitem{Lee2016}
N.~Lee, F.~Baccelli, and R.~W. Heath, ``Spectral efficiency scaling laws in dense random wireless networks with multiple receive antennas,'' \emph{{IEEE} Trans. Inf. Theory}, vol.~62, no.~3, pp. 1344--1359, Mar. 2016.

\bibitem{An2016}
K.~An, M.~Lin, J.~Ouyang, and W.-P. Zhu, ``Secure transmission in cognitive satellite terrestrial networks,'' \emph{{IEEE} J. Sel. Areas Commun.}, vol.~34, no.~11, pp. 3025--3037, Nov. 2016.

\bibitem{alzer1997}
H.~Alzer, ``On some inequalities for the incomplete gamma function,'' \emph{Mathematics of Computation}, vol.~66, no. 218, pp. 771--778, 1997.

\end{thebibliography}

\end{document}